\documentclass[12pt,epsfig]{article}

\usepackage{graphicx}
\usepackage{amsfonts}
\usepackage{amssymb}
\usepackage{amsmath}
\usepackage{mathdots}
\usepackage{placeins}
\usepackage{bbold}

\newskip\humongous \humongous=0pt plus 1000pt minus 1000pt
  \newif\ifdtup

\def\frac#1#2{ {{#1} \over {#2} }}

\def\eg{\hbox{\em e.g. }}
\def\ie{\hbox{\em i.e. }}

\def\beq{\begin{equation}}
\def\eeq{\end{equation}}

\def\beqn{\begin{eqnarray}}
\def\eeqn{\end{eqnarray}}

\def\de{\delta}

\let\de\partial
\let\mrm\mathrm
\let\mcal\mathcal

\let\mbb\mathbb

\newcommand{\bcas}{\begin{cases}}
\newcommand{\ecas}{\end{cases}}
\newcommand{\bmat}{\begin{pmatrix}}
\newcommand{\emat}{\end{pmatrix}}

\begin{document}

\title
{Thimble regularization at work:\\
 from toy models to chiral random matrix theories.}

\author
{F.~Di~Renzo and G.~Eruzzi\\
\small{Dipartimento di Fisica e Scienze della Terra, Universit\`a di Parma} \\
\small{and INFN, Gruppo Collegato di Parma} \\
\small{I-43100 Parma, Italy}\\
}

\maketitle

\begin{abstract}
We apply the Lefschetz thimble formulation of field theories to a 
couple of different problems. We first address the solution of a 
complex 0-dimensional $\phi^{4}$ theory. Although very simple, 
this toy-model makes us appreciate a few key issues of the
method. In particular, we will solve the model by a correct 
accounting of all the thimbles giving a contribution to the 
partition function and we will discuss a number of algorithmic 
solutions to simulate this (simple) model. We will then move to a 
chiral random matrix (CRM) theory. This is a somehow more realistic 
setting, giving us once again the chance to tackle the same couple of 
fundamental questions: how many thimbles contribute to the solution? how 
can we make sure that we correctly sample configurations on the thimble? 
Since the exact result is known for the observable we study (a condensate), 
we can verify that, in the region of parameters we studied, 
only one thimble contributes 
and that the algorithmic solution that we set up works well, despite
its very crude nature. The 
deviation of results from phase quenched ones highlights that in a 
certain region of parameter space there is a quite important sign 
problem. In view of this, the success of our thimble approach is quite 
a significant one.

\end{abstract}

\section{Introduction}

The so-called sign problem is one of the current big challenges for lattice 
field theories. It is in fact the major obstacle to tackling a non-perturbative 
study of the QCD phase diagram. Following pioneering work by Witten
\cite{Witten},  Lefschetz thimble regularization has been proposed as a
possible solution \cite{OurFirstTHMBL, Kikukawa} 
(for more recent contributions see also \cite{Yuya1, Yuya2, Yuya3, Yuya4}): 
the functional integral 
is defined in terms of fields taking values on non-trivial manifolds on
which the imaginary part of the action stays piecewise (\ie on the
distinct thimbles attached to different critical points) constant. It
is an elegant, although with many respects non-trivial alternative to
the standard formulation of field theories. It has intriguing
connections with resurgence theory, a few results of which motivate
the conjecture that {\em the semi-classical expansion of the path integral 
can be geometrized as a sum over Lefschetz thimbles}\footnote{This is
a literal citation from \cite{UnsalDorigoni}; we will have more to
quote later on the subject of resurgence.} \cite{UnsalDorigoni}.
Morse theory
\cite{MorseTHEORY} is the natural framework for discussing the thimble
regularization, even though it could be that it does not necessarily 
have the last word on the subject. 
In this work we will discuss the thimble solution of two different
models, having in mind two big issues. First of all, since there is 
a thimble attached to every critical point of the (complexified)
theory one is considering, we need to understand how many thimbles 
do give contribution to the solution of a given theory. A second
relevant problem is that of devising Monte Carlo algorithms to  
correctly sample thimbles, which are manifolds for which we lack 
a local description. \\

Toy models can be a precious tool to approach hard problems in
theoretical physics, with the hope that a simplified model can
nevertheless capture the relevant issues. Being the sign
problem both relevant and hard, it does not come as a surprise that
toy models have been around for a while. Quite interestingly, some of
these have been resisting the efforts to solve them in much the same 
way the real problem is still far for being fully solved. The first
application we discuss is the solution of a toy model that dates back to 
almost thirty years ago \cite{Ambjorn}. It can be regarded as a $0$-dim
version of a $\phi^4$ field theory. This simple model became a 
benchmark for the complex Langevin treatment of the sign problem, and 
quite interestingly only partial success has been claimed over the
years. We will show that the thimble regularization completely solves
the model. In this case we will show how in different regimes one or
more thimbles give contribution to the solution. Moreover, it is
possible to perform numerical simulations on the thimble(s): in this 
simple case we will have a number of algorithmic solutions at work. 
A partial account of these results has already been given in 
\cite{Lat2014phi40d}.\\

We will then address the solution of a chiral random matrix
theory. This is a somehow more realistic problem, for which once again
the application of the complex Langevin method has been shown to be 
non-trivial: in a given parametrization it fails \cite{KimRMT1}, while
in a different one it works \cite{KimRMT2}. Actually one can show that
the sign problem can be quite severe for this model. The theory has an
adjustable parameter (the dimension $N$ of the matrices) which
controls the dimensionality of the problem one has to solve. Since the
analytical solution for the observable we will study (a mass
condensate) is known, it appeared to us a perfect setting for testing
a conjecture: it could well be that more than one thimble contribute
(just like in the $0$-dim $\phi^4$ theory) in low dimensions, but as $N$
grows it could be that a single thimble dominates in the
thermodynamic limit. While we were ready for a richer scenario,
in the region of parameters we studied  
we actually did not find any other thimble but the one attached 
to the global minimum. We did not find any problem related to the 
parametrization of the theory; in particular, the parametrization 
that was failing in the case of complex Langevin 
was absolutely fine for the thimble treatment of the theory. 
In this case algorithmic problems are non-trivial. We will show how 
a natural parametrization of the thimble can be the
starting point for an algorithmic solution that in this case works in
its simplest version (admittedly a very crude one).\\

The paper is organized as follows. In section~2 we give a brief
account of the Lefschetz thimble approach to field theories: this is a
short review of results that are collected to facilitate the reader. 
Section~3 is dedicated to the $0$-dim $\phi^4$ toy model, showing that 
thimbles provide a complete solution; in particular we show that we
can effectively numerically simulate the model, making use of 
different algorithms. In section~4 we address the chiral random
matrix theory, showing that we can numerically solve it by thimble
regularization: all the analytical results are correctly
reconstructed. In the final section we draw a few conclusions and
mention natural steps forward.

\section{Thimble regularization in a nutshell}

In the following we collect the basics of thimble regularization: the
interested reader is referred to \cite{Witten,OurFirstTHMBL, Kikukawa}
for further details and references. \\

A conceptual starting point to approach the thimble regularization is
that of generalizing saddle point integration. The latter
displays a couple of features which appear as good candidates to
tackle the sign problem: stationary phase and localization. The full
generalization of saddle point techniques is formulated in the
framework of Morse theory \cite{MorseTHEORY}. 
\begin{itemize} 
\item One starts with an integral on a real domain of the form 
\begin{equation}
\int_{\mathcal{C}} d^nx \; g(x_1,\dots,x_n) \; \mbox{e}^{-S(x_1,\dots,x_n)}
\label{eq:MYintegral}
\end{equation}
in which $\mathcal{C}$ is a real domain of real dimension 
$n$\footnote{In what follows we will be a little bit sloppy in our
  notation: whenever there is not subscript attached to a symbol 
(e.g. $x$), that will denote a $n$-dimensional coordinate.} 
and both $S(x_1,\dots,x_n)=S_R(x_1,\dots,x_n)+iS_I(x_1,\dots,x_n)$ 
and $g(x_1,\dots,x_n)$ are holomorphic
functions. The notation for the exponential makes it clear that 
we have already in mind a functional integral (even if the normalizing
factor $Z^{-1}$ is missing). For such an
integral the following decomposition holds
\begin{equation}
\int_{\mathcal{C}} d^nx \; g(x_1,\dots,x_n) \; \mbox{e}^{-S(x_1,\dots,x_n)} = 
\sum_\sigma n_\sigma \; \int_{\mathcal{J}_\sigma} d^nz \; g(z_1,\dots,z_n) \; \mbox{e}^{-S(z_1,\dots,z_n)}
\label{eq:MORSE}
\end{equation}
in which an extension from a real domain to a complex one has been 
performed (see the complex variables $z_i=x_i+iy_i$ as opposed to the 
real ones $x_i$). (\ref{eq:MORSE}) holds in the homological sense, 
\ie $\mathcal{C} = \sum_\sigma n_\sigma \; \mathcal{J}_\sigma$.
\item The index $\sigma$ labels the critical points of the
  complex(ified) $S(z_1,\dots,z_n)$ and to each critical point $p_\sigma$ a stable
  thimble $\mathcal{J}_\sigma$ is attached. Each $\mathcal{J}_\sigma$
  is defined as the union of all the Steepest Ascent (in the following
  SA; we will also write SD for Steepest Descent) paths falling
  into $p_\sigma$ at (minus) infinite time, \ie the union of the
  solutions of 
\begin{eqnarray}\label{eq:thimblEQ}
\frac{dx_i}{d\tau} &=& \frac{\partial S_R(x,y)}{\partial x_i} \nonumber \\
\frac{dy_i}{d\tau} &=& \frac{\partial S_R(x,y)}{\partial y_i} 
\end{eqnarray}
satisfying $z(\tau=-\infty) = x(\tau=-\infty) + iy(\tau=-\infty) =
p_\sigma$. The real dimension of each thimble $\mathcal{J}_\sigma$ is
  $n$. It is quite natural to regard thimbles as manifolds embedded in 
$\mathbb{C}^n$ (which is instead of real dimension $2n$).
\item For each critical point one also defines unstable thimbles 
$\mathcal{K}_\sigma$ as the union of all flows satisfying 
equation (\ref{eq:thimblEQ}) and going to the critical point in the
opposite time limit, \ie such that 
$z(\tau=\infty) = p_\sigma$. The coefficients
$n_\sigma$ count the intersections of the $\mathcal{K}_\sigma$ with 
the original domain of integration 
$n_\sigma = \left< \mathcal{C}, \mathcal{K}_\sigma\right>$. 
\item The imaginary part $S_I$ stays constant on thimbles, \ie there
  is a phase associated to each thimble.
\end{itemize} 
Note that in the framework of field theories a natural picture 
of universality emerges. A single thimble can give us 
a formulation of a field theory with the same degrees of freedom, 
the same symmetries\footnote{Since we have discussed the case of 
a non-degenerate hessian, one could wonder how the method can be
  applied in case where Spontaneous Symmetry Breaking (SSB) is in
  place. In \cite{BoseTaming} a solution has been described and shown
  to be effective: one introduces an explicit Symmetry Breaking term
  $h$ and studies the limit $h \rightarrow 0$. This is not the only
  way to proceed: for a different thimble approach to SSB the reader 
is referred to \cite{Kikukawa}. Symmetries are dealt in yet another 
way in the case of Gauge Theories: the construction of thimbles
was discussed in \cite{AtBott} and reviewed in \cite{Witten};  
see also the discussion in \cite{OurFirstTHMBL,LAT2015_gauge}.}
 and symmetry representations, the same 
Perturbation Theory and naive continuum limit of the original formulation
(see \cite{OurFirstTHMBL} for details). In force of universality
we expect that these properties essentially determine the behavior of
physical quantities in the continuum limit. 
Moreover a simple argument suggests that in the thermodynamic limit
only thimbles attached to global minima can survive, as it is easily 
seen (we now call $\phi_\sigma$ the critical points, having in mind
field configurations, and consider the partition function of the field
theory) 
$$
Z = \sum_\sigma n_\sigma \; \mbox{e}^{-iS_I(\phi_\sigma)}
\int_{\mathcal{J}_\sigma} d^nz \; \mbox{e}^{-S_R(z)} = 
\sum_\sigma n_\sigma \; \mbox{e}^{-S(\phi_\sigma)}
\int_{\mathcal{J}_\sigma} d^nz \; \mbox{e}^{-(S_R(z)-S_R(\phi_\sigma))}
$$
In the end, it could well be that a full resolution in terms of all 
the thimbles could turn out to be with many respect overkilling the
original problem. Having said this, we nevertheless stress that both 
the universality argument and the
thermodynamic argument can not be regarded as conclusive. It
  is worth noting that resurgence theory \cite{MU} with many respect 
even {\em asks for}
  more than one thimble in view of the interpretation of the
  semi-classical approximations as trans-series\footnote{A nice account of
  many issues connected to resurgence has been recently provided in 
\cite{MUlat2015}.}. All in all, the fact that in certain cases a 
single thimble dominance can take place has to be regarded as a
conjecture: as we will see, this was in a sense 
one of the motivation of this work on a Random Matrix Model. The
subject deserves deeper investigation and we think it will certainly
receive it. \\

It is good to have a somehow more constructive approach to the thimble
formulation. We therefore now sketch a few more technical details
that the reader will see at work in the following sections. The
integral we have in mind will be the functional integral of a field
theory. 
Let us first of all parametrize the field in the vicinity of a critical
point as $\Phi_i = \phi_i-\phi_{\sigma,i}$. Here and in the 
following $i$ is a multi-index; in particular it can refer to a real
or imaginary part. The real part of the action can be expressed as 
\beq
S_R\left(\phi\right)=S_R\left(\phi_\sigma\right)+\frac{1}{2}\Phi^TH\Phi+\mcal{O}\left(\phi^3\right) 
\label{eq:ACTexpansion}
\eeq
where the $2n\times 2n$ matrix $H$ is the hessian evaluated at the critical
point 
$$
H_{ij}=\frac{\de^2S_R}{\de\phi_i\de\phi_j}\biggl|_{\phi=\phi_\sigma}
$$
H can be put in diagonal form 
$$
\Lambda=\mrm{diag}\left(\lambda_1,\cdots,\lambda_n,-\lambda_1,\cdots,-\lambda_n\right)
$$
by a transformation $H=W\Lambda W^T$ defined by the orthogonal
matrix $W$ whose columns are given by the normalized
eigenvectors of $H$, that is $\{v^{\left(i\right)}\}_{i=1\cdots 2n}$. Half of
the eigenvalues of $H$ are positive; the corresponding eigenvectors
span the tangent space to the thimble at the critical point. 
Any combination of these vectors is a direction along which the real part
of the action grows. If we leave the critical point along these
directions integrating the SA equations we span the
thimble. On the
other side, the other directions (which are attached to the negative
eigenvalues) would take us along the unstable thimble. \\
At a generic point $Z\in\mcal{J}_\sigma$ we miss a priori the
knowledge of the tangent space $T_Z\mcal{J}_\sigma$; in general we
expect that the latter is not parallel to the canonical basis 
of $\mbb{C}^n$ whose duals appear in    
$\mrm{d}^n z=\mrm{d}z_1\wedge\cdots\wedge\mrm{d}z_n$. We thus want to
perform the relevant change of coordinates from the canonical ones 
(of $\mbb{C}^n$) to the basis of $T_Z\mcal{J}_\sigma$, given by the 
(complex) vectors $\{U^{\left(i\right)}\}_{i=1\cdots n}$ (these are
orthonormal with respect to the standard hermitian metric of
$\mbb{C}^n$). 
Let $\varphi:N\subset\mcal{J}_\sigma\rightarrow\mbb{R}^n$ 
be a local chart in a neighbourhood $N\subset\mcal{J}_\sigma$ of $Z$
$$
\varphi\left(Z+\sum_{i=1}^n U^{\left(i\right)}y_i\right)=Y+\mcal{O}\left(y^2\right)\in\mbb{R}^n
$$
If we denote $U$ the $n\times n$ complex unitary matrix whose columns
are the vectors $\{U^{\left(i\right)}\}$, we can express the integral of a
generic function $f\left(Z\right)$ on the thimble as
\beq
\int\limits_N\mrm{d}^n z\,f\left(Z\right)=
\int\limits_{\varphi\left(N\right)}\prod_{i=1}^n\mrm{d}y_i\,f\left(\varphi^{-1}\left(Y\right)\right)
\,\det{U}\left(\varphi^{-1}\left(Y\right)\right)
\label{eq:RESphase}
\eeq
In this expression the quantity $\det{U}=e^{i\omega}$ ($U$ is unitary)
has appeared; this is what has been termed the residual phase 
\cite{OurFirstTHMBL} (see \cite{resPHASE} for further details). 
This could in principle re-introduce a sign
problem in the thimble formulation, but it is expected that this is
not the case. Not any phase gives raise to a serious sign problem, 
and in particular one expects that a phase changing rather smoothly 
can be safely taken into account by reweighting. This
expectation could appear optimistic, but has been till now confirmed
(see \cite{Kikukawa}) and will be confirmed also in this work. \\

We end this brief introduction to the thimble formulation by
going back to the constructive point of view: we can span the thimble
by integrating the SA equations for the field $\phi$~\footnote{It is 
worth to recall here that the subscript $i$ is a multi-index, in which
real and imaginary parts are on the same footing.}
$$
\frac{\mrm{d}\phi_i}{\mrm{d}t}=\frac{\de S_R}{\de\phi_i}\qquad i=1\cdots 2n
$$
This has a counterpart in parallel-transport equations for the 
$n$ basis vectors which defines the tangent space to the thimble 
(see \cite{OurFirstTHMBL,Kikukawa})
\beq
\frac{\mrm{d}V^{\left(i\right)}_j}{\mrm{d}t}=\sum_{k=1}^{2n}
\frac{\de^2 S_R}{\de\phi_k\de\phi_j}V^{\left(i\right)}_k
\qquad i=1\cdots n\quad j=1\cdots 2n.
\label{eq:VecTransport}
\eeq
We can set up a similar equation for any other vector 
with an initial condition on the tangent space at the critical point; 
(\ref{eq:VecTransport}) expresses the parallel transport of a vector
along the gradient flow. 
In the vicinity of the critical point one knows the 
asymptotic ($t\rightarrow-\infty$) solutions 
\begin{equation}
t\ll 1 \;
\bcas
\; \phi_j\left(t\right)\approx\phi_{\sigma,j}+\sum\limits_{i=1}^n v^{\left(i\right)}_j e^{\lambda_i t}n_i & j=1\cdots 2n\quad\left|\vec{n}\right|^2=1\\
\; V^{\left(i\right)}_j\left(t\right)\approx v^{\left(i\right)}_j e^{\lambda_i t} & j=1\cdots 2n\quad i=1\cdots n
\ecas
\label{eq:asymptSOL}
\end{equation}
Note that from a practical point of view the former parametrization 
is viable only provided one introduces a reference time $t_0\ll 1$ at
which the former asymptotic solution holds.
We will make estensive use of the former equations, which in
particular can be regarded as initial conditions for a given flow on
the thimble, \eg for eq.~(\ref{eq:VecTransport}). 
It thus emerges a natural picture in which 
a generic point $\Phi\in\mcal{J}_\sigma$ is unambiguously defined by a 
choice of $\hat{n}$\footnote{The hat notation remembers us of the
normalization condition, \ie $\hat{n}$ singles out a direction in the
tangent space at the critical point.} and the time 
$t$: $\Phi=\Phi\left(\hat{n},t\right)$ 
(this has been very effectively discussed in \cite{Kikukawa}). Note also that one could
insist on regarding the (\ref{eq:asymptSOL}) as valid all over; this
would in turn mean one is considering a purely quadratic action
(\ie the free field approximation).

\section{The $0$-dim $\phi^4$ toy model} 

We will now put at work what we have just seen, applying the thimble
regularization to the study of the action
$$
S\left(\phi\right)=\frac{1}{2}\sigma\phi^2+\frac{1}{4}\lambda\phi^4
$$
\noindent with $\phi\in\mathbb{R}$, 
$\lambda\in\mathbb{R}^{+}$ and
$\sigma=\sigma_{R}+i\,\sigma_{I}\in\mathbb{C}$. 
This is obviously a toy model, and one regards as 
correlators plain one-dimensional integrals such as
\begin{equation}
\label{eq:moments}
\left<\phi^n\right>=\frac{1}{Z}\int\limits_\mathbb{R} \mathrm{d}\phi\,
\phi^n \, e^{-S\left(\phi\right)} 
\end{equation}
\noindent with the partition function given by:
$$
Z=\int\limits_\mathbb{R} \mathrm{d}\phi\, e^{-S\left(\phi\right)}
$$
The solution is given in terms of a modified Bessel function, 
\ie $Z=\sqrt \frac{\sigma}{2\lambda} \,
e^{\frac{\sigma^2}{8\lambda}}
\, K_{-\frac{1}{4}}(\frac{\sigma^2}{8\lambda}) $, differentiating 
appropriately which one can get any of the (\ref{eq:moments}).
The choice of a complex $\sigma$ is a prototypal case
of the sign problem: with a complex action, we miss a positive
semi-definite measure and hence a probability distribution to start 
with; in particular, a direct access to Monte Carlo methods
is ruled out. \\

It was realized long time ago that a solution to the sign problem
could be searched in the context of Stochastic Quantization: the Langevin
equation admits a formal solution also for complex actions, in
particular via the Fokker-Planck formulation \cite{Parisi,Klauder}. 
Turning the formal arguments into a rigorous proof eventually turned
out to be hard and numerical
instabilities (suggesting problems) were in particular discussed in the context
of the theory at hand \cite{Ambjorn}. Much experience has been gained
over the years and much progress has been done \cite{CLang1}. 
The question of
convergence of complex Langevin equation remains a subtle one, and
quite interestingly even the simple model at hand displays delicate issues. 
For a recent and thorough study of complex Langevin dynamics of
this model, the reader can refer to 
\cite{Gert_phi4_0d}. 
One peculiar feature of this model is that complex Langevin
simulations display divergences for $\left<\phi^n\right>$ 
with $n>4$ in a certain region of parameters. The relation between
complex Langevin and thimbles has been investigated in 
\cite{Gert_THIMBLEphi4_0d,SeilerGert_THIMBLE}.\\

We can complexify the field by setting $\phi=x+i\, y$. As a result,
real and imaginary part of the action read
\begin{eqnarray*}
S^{R}&=&\frac{1}{2}\left[\sigma_{R}\left(x^{2}-y^{2}\right)-2\sigma_{I}xy\right]+\frac{1}{4}\lambda\left(x^{4}+y^{4}-6x^{2}y^{2}\right)\\
S^{I}&=&\frac{1}{2}\left[\sigma_{I}\left(x^{2}-y^{2}\right)+2\sigma_{R}xy\right]+\lambda\left(x^{3}y-xy^{3}\right)
\end{eqnarray*}

The hessian is built from the second derivatives of $S_R$ and takes the form:

\begin{equation}
H\left(x,y\right)=\begin{pmatrix}\sigma_{R}+3\lambda x^{2}-3\lambda y^{2} & -\sigma_{I}-6\lambda xy\\
-\sigma_{I}-6\lambda xy & -\sigma_{R}-3\lambda x^{2}+3\lambda y^{2}
\end{pmatrix}
\label{eq:hessian0d}
\end{equation}

There are 3 critical points: $\phi_{0}=0$ and $\phi_{\pm}=\pm\sqrt{-\frac{\sigma}{\lambda}}$
(which are the two, complex valued, ``Higgs vacua''). The question is
now which thimbles do give a contribution to the integrals we want to
compute, and the answer is quite different in the 3 cases 
$\sigma_{R}>0$, $\sigma_{R}<0$
and $\sigma_{R}=0$: in each case we computed the stable and unstable
thimbles associated to each critical point. This can be done putting
at work the constructive definition of thimbles we discussed in
the previous section.

\begin{figure}[ht]
\begin{center}
  \begin{tabular}{cc} 
 \hspace{-1.cm}
     \includegraphics[height=6.8cm,clip=true]{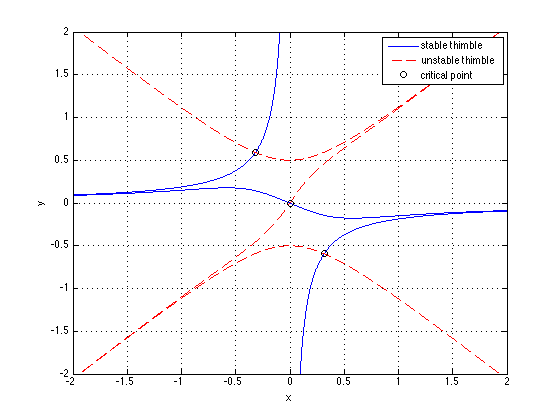}
     &
 \hspace{-1.cm}
     \includegraphics[height=6.8cm,clip=true]{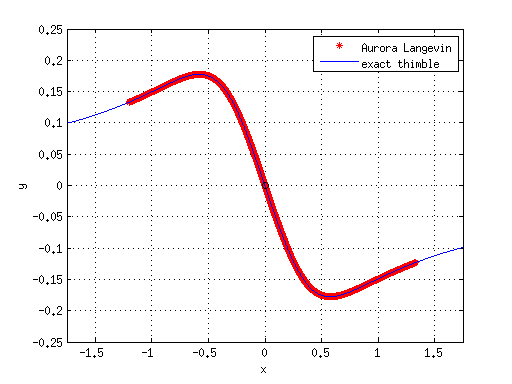}
  \end{tabular}
\end{center}
  \caption{Thimbles structure for $\sigma = 0.5 + i 0.75$, 
$\lambda = 2$ (left panel). In this case only the unstable thimble
attached to $z = 0$ intersects the real axis and thus only one
critical point contributes. On the right we can see how the Langevin
simulation correctly covers the relevant thimble.}
  \label{fig:phi40d_pos}
\end{figure}

\begin{figure}[ht]
\begin{center}
  \begin{tabular}{cc} 
 \hspace{-1.cm}
     \includegraphics[height=6.8cm,clip=true]{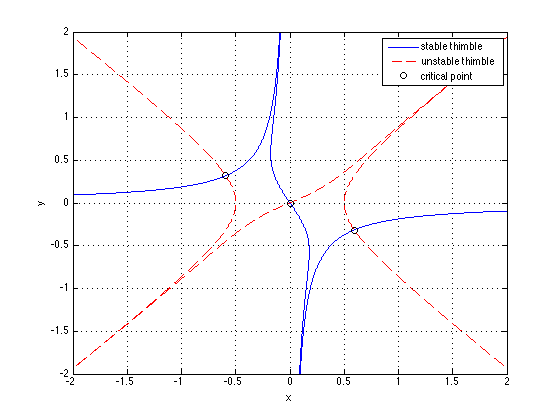}
     &
 \hspace{-1.cm}
     \includegraphics[height=6.8cm,clip=true]{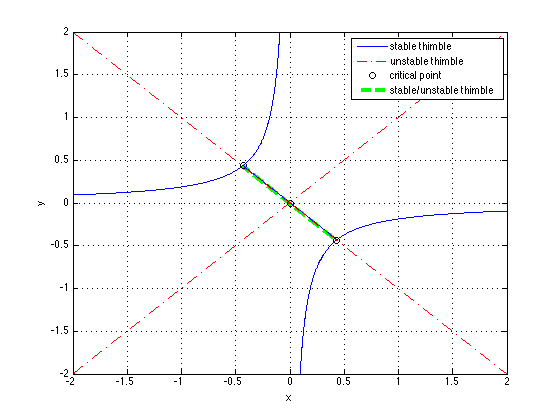}
  \end{tabular}
\end{center}
  \caption{Thimbles structure for $\sigma = -0.5 + i 0.75$, 
(left panel) and $\sigma = i 0.75$ (right); in both cases 
$\lambda = 2$. For $\sigma_R < 0$ (left) all the three 
critical points contribute. $\sigma_R = 0$ (right) is an example of a
Stokes phenomenon.}
  \label{fig:phi40d_neg_stokes}
\end{figure}

In practice, we want to integrate the equations of SA 
starting in the vicinity of the critical point $\phi_\sigma$ for an 
arbitrarily long flow time $t$. We can do this provided that the 
initial condition is chosen correctly: for the stable thimble 
this means we leave the
critical point along the direction (in the $xy$ plane) which is 
given by the eigenvector of positive eigenvalue of the hessian 
(\ref{eq:hessian0d}) computed at the critical point. Once
we have singled out the relevant direction, we can ascend in two ways
(namely, increasing or decreasing $x$), both of which we have to take
to cover all the thimble. By holomorphicity 
the hessian has two eigenvalues opposite in sign. Since $S_R$ 
always increases along the flow, $\exp{\left(-S_R\right)}$ goes to $0$ 
as $t\rightarrow+\infty$, thus ensuring convergence of the integrals 
along the thimble. To obtain the unstable thimble
$\mathcal{K}_\sigma$, we can repeat the same procedure described
above, but picking up the eigenvector of the hessian of $S_R$ with
negative eigenvalue. Note that the unstable thimble is needed because 
the coefficient $n_\sigma$ in our master equation (\ref{eq:MORSE})
counts the intersection of such thimbles with the original domain of 
integration, which in our case is the real axis (the sign ambiguity 
is not resolved just by this definition, but it can be deduced by 
means of other considerations). 
Figures \ref{fig:phi40d_pos} (left panel) and 
\ref{fig:phi40d_neg_stokes} show the results for the three cases
$\sigma_{R}>0$, $\sigma_{R}<0$ and $\sigma_{R}=0$ 
(see also \cite{OURmovie}).

From figure \ref{fig:phi40d_pos}, we see that when $\sigma_{R}>0$ 
the unstable thimbles related to the Higgs vacua do not intersect 
the real axis. 
Therefore these points do not contribute to the integrals, that is 
$n_{\pm}=0$ and $n_{0}=1$. By integrating along the stable thimble 
attached to $\phi_0$, we recover the correct results for, say, 
$Z=\int e^{-S}$ (the integration can be easily carried on along 
the real axis both analytically and numerically). The case 
$\sigma_{R}<0$ depicted in the left panel of figure
\ref{fig:phi40d_neg_stokes} is a totally different matter, 
as we cross the Stokes ray $\sigma_{R}=0$ while changing sign 
to $\sigma_R$. Now we see that the unstable thimbles connected 
to the Higgs vacua do intersect the real axis and therefore 
$n_{\pm}\neq0$, as well as $n_0\neq0$. The correct combination 
which recovers the expected results for the integrals turns 
out to be $n_{0}=-1$ and $n_{\pm}=+1$. What is the origin of 
this discontinuity? and, above all, if we hadn't known 
the correct result from 
the beginning, how would have we calculated the $n_{\sigma}$? 
The answer lies in considering the case $\sigma_{R}=0$, 
showed in the right panel of 
figure~\ref{fig:phi40d_neg_stokes}. The stable thimble connected 
to $0$ exhibits the Stokes phenomenon: in fact it ``collapses'' 
into the Higgs vacua, from which it does not ``move'' any more; 
the unstable thimble continues to say that $n_{0}\neq0$. 
The stable thimbles connected to the Higgs vacua display the same 
shape, but their unstable counterparts collapse into $0$ 
(by overlapping its stable thimble) and therefore there is 
intersection with the real axis; so, $n_{\pm}\neq0$. However, 
there is no integer-valued combination of $n_{\sigma}$ that recovers 
the correct results for $\sigma_{R}=0$. This is quite expected, 
as the Morse decomposition along thimbles is not legitimate when 
we are on a Stokes ray, on which we clearly are (the imaginary axis 
in the complex $\sigma$ plane is a 
``Stokes ray'')\footnote{see \cite{Witten} for a detailed explanation 
of the Stokes phenomenon with respect to the Airy integral}. 
Now, the original integral is
continuous (in fact, it is holomorphic) in $\sigma$ and therefore 
there cannot be any discontinuity in the computation of the 
partition function $Z$ in $\sigma_{R}=0$. Thus, we must have 
$Z\left[\sigma_{R}\rightarrow0^{+}\right]=
Z\left[\sigma_{R}\rightarrow0^{-}\right]=
Z\left[\sigma_{R}=0\right]$. 
By examining the integration along the thimble connected to $0$, 
we find that it is discontinuous in $\sigma_{R}=0$, and again, 
this is not surprising as the thimble shape undergoes a radical 
change between the two cases. The change in sign of the 
$n_{\sigma}$ is precisely the only one which keeps the original 
integral continuous while crossing the Stokes ray.

\subsection{A variety of algorithmic solutions}

Within the thimble regularization we were able to perform numerical 
simulations of the quartic toy model, making use of different
algorithms. In particular, we were able to numerical compute all the
possible moments (\ref{eq:moments}). \\

It was observed in \cite{OurFirstTHMBL} that the Langevin algorithm 
is the obvious candidate for sampling configurations on the thimble. 
In
\beq
\frac{\mrm{d}\phi_i}{\mrm{d}t}=- \frac{\de S_R}{\de\phi_i} + \eta_i \qquad i=1\cdots 2n
\eeq
the drift term constrains the field on the thimble by definition, so
that the problem boils down to extracting a convenient noise, \ie a
noise tangent to the thimble. We do not discuss here the original
solution which was put forward in \cite{OurFirstTHMBL} (the Aurora 
algorithm); there will be a convenient time for such a discussion 
when we later approach the CRM model. Here it suffices to say that, 
being the thimble $1$-dimensional, at every point the tangent space 
reduces to the direction singled out by the drift term itself. As a
matter of fact, Langevin works pretty well; in the right panel of 
figure \ref{fig:phi40d_pos} one can see how the simulation 
correctly samples configurations on the thimble. Here parameters are 
the same of the left panel, so one thimble is relevant, \ie the one 
attached to the 
origin (which in the notation of (\ref{eq:MORSE}) we denote 
$p_0$). The (Aurora) Langevin algorithm samples points according to 
the measure normalized by\footnote{$\tau$ is the real coordinate on 
the (one-dimensional) thimble.}
\beq
\mcal{Z}^{\left(0\right)} \equiv 
\int_{\mathcal{J}_{0}}\mrm{d}\tau\,e^{-S_R}
\label{eq:Zs0}
\eeq
We now denote 
\beq
\langle O \rangle_{0} \equiv
\frac{\int_{\mathcal{J}_{0}}\mrm{d}\tau\,O\,e^{-S_R}}
{\mcal{Z}^{\left(0\right)}}
\eeq
and stress that this is not what we have to compute. Properly 
including the residual phase, the correct result was computed as
\beq
\langle O \rangle = 
\frac{\langle e^{i\omega} O \rangle_{0}}{\langle e^{i\omega} \rangle_{0}}
\label{eq:VEVs0}
\eeq

When $\sigma_R<0$ the thimbles associated to all the three critical 
points\footnote{In the notation of (\ref{eq:MORSE}) we now denote 
$\phi_+=p_1$ and $\phi_-=p_2$.} contribute and we have to compute

\begin{equation} 
\langle O \rangle = \frac{\sum_{i=0}^{2} n_i \,
  e^{-i\,S_I\left(p_{i}\right)} \int_{\mathcal{J}_{i}}
  \mrm{d}\tau\, e^{-S_R} \,O\, e^{i\omega}}{\sum_{i=0}^{2} n_i \,
  e^{-i\,S_I\left(p_{i}\right)} \int_{\mathcal{J}_{i}}
  \mrm{d}\tau\, e^{-S_R}  \, e^{i\omega}}
\label{eq:MANYthimbles}
\end{equation}

which can be written

\begin{equation} 
\langle O \rangle = 
\frac{\langle e^{i\omega} O \rangle_{0} + \alpha_1 \langle e^{i\omega}
  O \rangle_{1} + \alpha_2 \langle e^{i\omega} O \rangle_{2}}
{\langle e^{i\omega}\rangle_{0} + \alpha_1 \langle e^{i\omega}
  \rangle_{1} + \alpha_2 \langle e^{i\omega} \rangle_{2}}
\end{equation}

with

\begin{equation} 
\alpha_i = \frac{n_i \, e^{-i\,S_I\left(p_{i}\right)} \mcal{Z}^{\left(i\right)}}
{n_0 \, e^{-i\,S_I\left(p_{0}\right)} \mcal{Z}^{\left(0\right)}}
\;\;\;\;\; i=1,2
\end{equation}

On each thimble ${\mathcal{J}_{i}}$ ($i=0,1,2$) the quantities 
$\langle e^{i\omega} O \rangle_{i}$ and $\langle e^{i\omega} \rangle_{i}$
can be computed via (Aurora) Langevin simulations. The (complex) 
unknown coefficients $\alpha_i$ can then be fixed by relations which
can be regarded as {\em renormalization conditions in a physical
  scheme}, \ie

\begin{equation}
\frac{\langle e^{i\omega} O_i \rangle_{0} + \alpha_1 \langle e^{i\omega}
  O_i \rangle_{1} + \alpha_2 \langle e^{i\omega} O_i \rangle_{2}}
{\langle e^{i\omega}\rangle_{0} + \alpha_1 \langle e^{i\omega}
  \rangle_{1} + \alpha_2 \langle e^{i\omega} \rangle_{2}} 
= X_i \;\;\;\;\; i=1,2
\end{equation}

where the $X_i$ are known values of given observables $O_i$ (\eg, in the case of moments 
(\ref{eq:moments}), two of them). As always in such an approach, 
one gives up predicting everything, but after normalizing results to a
(minimum) number of external inputs, one has full predictive power for
(all the) other quantities. Of course computing the moments 
(\ref{eq:moments}) for the toy model at hand is not such a big 
numerical success; nevertheless the outline of the method is quite 
general. In particular, we will refer to it in section 4.3. \\

Another algorithmic solution for this simple setting is provided by the
Metropolis algorithm which is described in \cite{AbiMarcoGigi}. The 
method relies on a correspondence between the full model one has to
simulate and a gaussian approximation associated to it. The latter is 
obtained by diagonalizing the hessian at a critical point and
truncating the expansion of the action around it, \ie

\beq
S_R(\eta) = S_R(\phi_\sigma) + \frac{1}{2} \sum_{k=1}^{2n} \lambda_k
\eta_k^2 \equiv S_R(\phi_\sigma) + S_G(\eta)
\label{eq:GaussianAction}
\eeq

where the $\eta_k$ are the $\Phi_k=\phi_k-\phi_{\sigma,k}$ of equation 
(\ref{eq:ACTexpansion}) expressed in the basis provided by the eigenvectors of
the hessian, with a convenient ordering in which $\lambda_k>0$ for 
$k=1 \ldots n$ and $\lambda_k<0$ for $k=n+1 \ldots 2n$. 
For the gaussian action (\ref{eq:GaussianAction}) it
is very simple to construct the associated stable thimble. It is a
{\em flat} thimble in which the tangent space is known once and
for all, \ie the span of the eigenvectors associated to 
$\left\{\lambda_k|k=1 \ldots n\right\}$: we term it a {\em gaussian thimble}.
For the gaussian thimble the solution in the right hand side of 
(\ref{eq:asymptSOL}) is valid all over the manifold. \\

The simulation is run as a quite 
standard Metropolis algorithm controlled by an accept/reject test, 
with a mechanism for proposing configurations which is dictated by the
correspondence between the thimble one has to sample and its 
gaussian approximation. 
We sketch the method in the case of more than one thimble 
contributing to the final result, to stress how also in this case
we were able to run numerical simulations on thimbles, for both 
$\sigma_R>0$ and $\sigma_R<0$. 

The method always handles a couple of configurations, \ie one $\phi$ 
field on the thimble we have to sample and one auxiliary $\eta$ field 
on the associated gaussian thimble. In order
to extract a new $\phi'$ field one proceeds as follows:
\begin{itemize}
\item One proposes a thimble $\sigma'$ (\ie a critical point) with a probability
$$
\frac{|n_{\sigma'}|}{\sum_\sigma |n_\sigma|}
$$
\item One extracts a configuration $\eta'$ on the gaussian thimble associated
  to that critical point according to the weight $e^{-S_G}$. 
This is trivial, given the gaussian form.
\item One starts a SD on the gaussian thimble with
  $\eta'$ as initial condition. The integration is carried on over a
  time extent $\bar \tau$ such that one ends up close enough to the critical
  point, namely at a point where the gaussian thimble and the thimble
  one has to sample
  effectively sit on top of each other (this means that the solution 
(\ref{eq:asymptSOL}) holds for both thimbles). We call $\bar \eta$ the
configuration that has been obtained in this way.
\item Taking $\bar \eta$ as the initial condition, one integrates 
the SA equations for the complete theory over the same 
time extent $\bar \tau$. This generates the new configuration $\phi'$.
\item $\phi'$ is accepted with probability
$$
P_{acc} =
\mbox{min} \left\{ 1, \,e^{-[S_R(\phi')-S_R(\phi)]+[S_G(\eta')-S_G(\eta)]} \right\}
$$
\end{itemize}

The result for a given observable $O$ is obtained as
$$
\langle O \rangle = 
\frac{\frac{1}{T} \sum_{t=1}^{T} e^{i \omega(\phi_t)} O(\phi_t)}
{\frac{1}{T} \sum_{t=1}^{T} e^{i \omega(\phi_t)}}
$$

where the index $t$ runs over all the configurations sampled by
Metropolis. \\

\begin{figure}[t]
\begin{center}
  \begin{tabular}{cc} 
 \hspace{-1.cm}
     \includegraphics[height=6.8cm,clip=true]{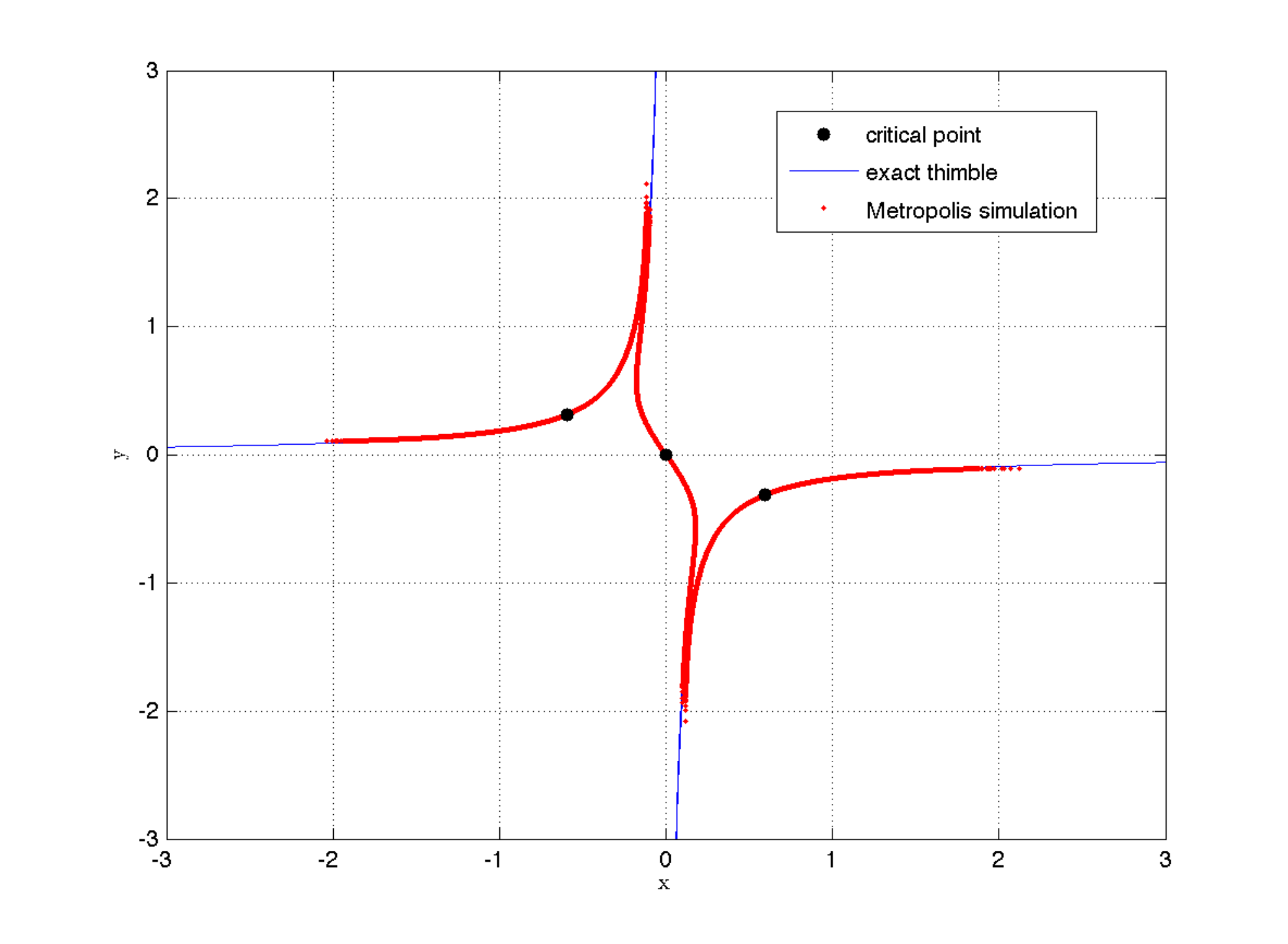}
     &
 \hspace{-1.cm}
     \includegraphics[height=6.8cm,clip=true]{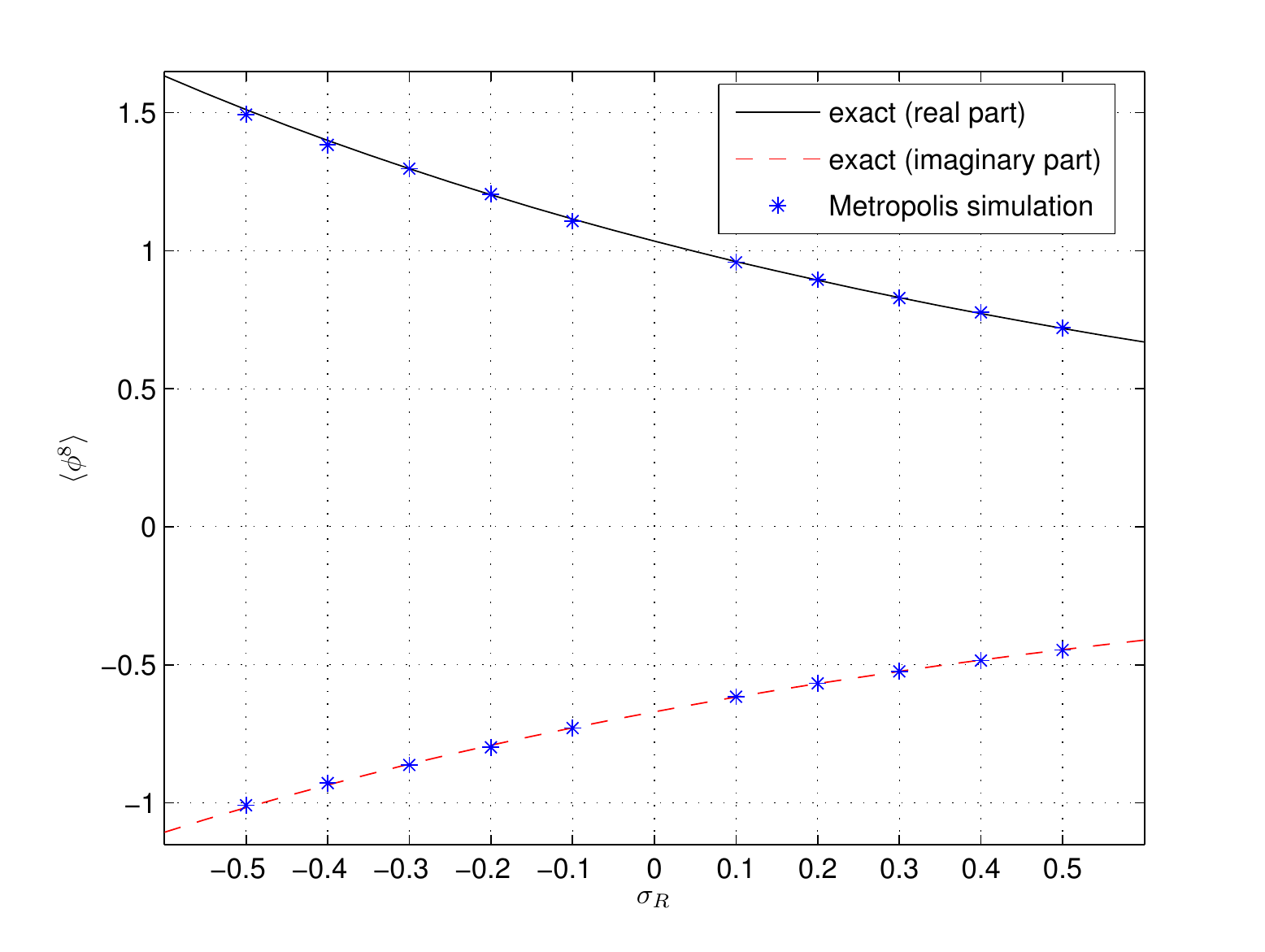}
  \end{tabular}
\end{center}
  \caption{Left panel: the three thimbles associated to 
$\sigma = -0.5 + i 0.75$ correctly sampled by a Metropolis 
simulation. Right panel: for the same choice of parameters, the 
computed values of $\langle \phi^8 \rangle$ over a range
of both $\sigma_R>0$ and $\sigma_R<0$.}
  \label{fig:phi40d_Metropolis}
\end{figure}

Notice that the previous accounting of the Metropolis algorithm is
technically different from the proposal of 
\cite{AbiMarcoGigi}\footnote{We decided to enlighten the rationale of
  the algorithm, leaving out the technicalities.}. The
latter relies on an exponential mapping for the integration time (\ie 
$r=e^{-t}$) and an adjustable parameter is introduced to control 
convergence properties (the interested reader can refer to figure 4 
of \cite{Lat2014phi40d}). For a given, effective choice of this
parameter figure \ref{fig:phi40d_Metropolis} displays how
the three thimbles giving contribution in the region $\sigma_R<0$
are sampled in a Metropolis simulation. \\

We stress that also in this
case one could think of situations in which the weights $n_\sigma$ are
unknown. However, here the situation is different from that of
Langevin, since we only need to know a few integers values. In other
terms, given the knowledge of the set of relevant integers
$\{n_\sigma\}$, 
the problem is
solved on an entire region of parameter space: in the case at hand,
for each $\sigma_R<0$ (technically, over the entire region which ends
up in a point where a Stokes phenomenon shows up). Notice that in
principle there can be different ways of finding the relevant set of
integers (\eg known asymptotic solutions in a convenient 
region)\footnote{The $\alpha_i$ introduced for the (Aurora) Langevin
  algorithm entail instead the values of partition functions and are
  given at a given point of parameter space; in the case at hand, for
  a given value of $\sigma_R<0$.}.

Both (Aurora) Langevin and Metropolis could correctly compute the
moments (\ref{eq:moments}), in both regions $\sigma_R>0$ and 
$\sigma_R<0$. For example, figure \ref{fig:phi40d_Metropolis} (right panel)
displays the computed values of $\langle \phi^8 \rangle$ over a range
of both $\sigma_R>0$ (one thimble being relevant) and $\sigma_R<0$ 
(three thimbles to be taken into account) values. \\

Note that there is another natural way of computing on a thimble,
and this takes advantage of the fact that on a thimble there is a one-to-one
correspondence in between configurations and values of $S_R$. 
To make things simple, let us consider the case in which only one
thimble is relevant and let us write the partition function 
$Z=Z_{\mbox{{\tiny up}}}+Z_{\mbox{{\tiny down}}}$: these are the two contributions
resulting from the two pieces of the thimble we have already referred to.
Namely, they are associated to leaving the 
critical point along the direction dictated by the eigenvector of the
hessian in one of the two possible ways (\ie increasing or decreasing
$x$ values). Each $Z_{\mbox{{\tiny up/down}}}$ has the global phase $e^{-i\,S_I\left(p_\sigma\right)}$ 
as a factor and features an integrand which is the residual phase 
times a monotonic function of $S_R$. It thus can be written 
taking the action as the integration variable, \eg 
\begin{equation} 
Z_{\mbox{{\tiny up}}} = e^{-i\,S_I\left(p_\sigma\right)}
\int_{S_{p_\sigma}}^\infty dS \; 
\mbox{e}^{-S_R} \, | \nabla S_R |^{-1}\,e^{i\,\tan^{-1}\left(\de_yS_R/\de_xS_R\right)}
\label{eq:Zi}
\end{equation} 
We could have written the integral by taking the
flow time as the integration variable (also in this case there is a 
one-to-one correspondence with the configurations along the thimble,
each reached at a given flow time). Note that in computing
(\ref{eq:Zi}) one proceeds by integrating the SA. We
illustrated the issue by taking into account the $Z$, but we
showed that all the moments (\ref{eq:moments}) can be successfully
computed in this way. In a sense, (\ref{eq:Zi}) is the
prototype of a parametrization we will see at work for the CRM model. \\
 
All in all, we think that the simple toy model we discussed is a
perfect playground to see thimble regularization at work: it is 
instructive both from the point of view of inspecting the structure 
of relevant thimbles and from the algorithmic point of view (we can
compute on thimbles).

\section{Chiral random matrix model}

\begin{figure}[t]
\begin{center}
\includegraphics[height=9cm,clip=true]{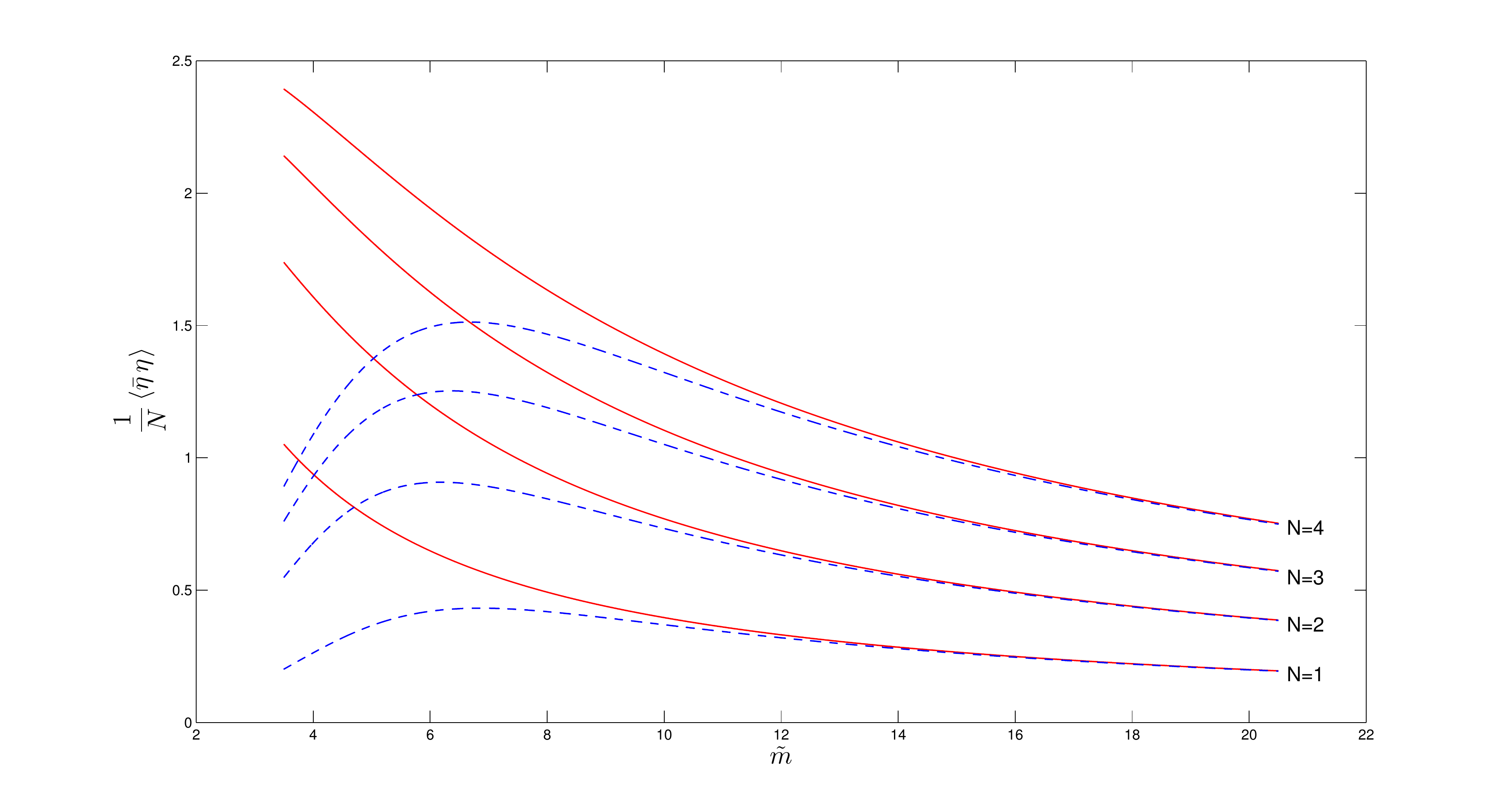}
\caption{Exact (solid red line) and phase quenched 
(dashed blue line) results for the condensate, 
at fixed $N_f=2$, $\tilde{\mu}=2$.}
\label{fig:PhaseQuenchedFailure}
\end{center}
\end{figure}

We now address the chiral random matrix model 
defined by the partition function
\begin{eqnarray}
Z_{N}^{N_{f}}(m) & = & \int d\Phi d\Psi\:\mbox{det}^{N_{f}}\left(D(\mu)+m\right)\exp\left(-N\cdot\mbox{Tr}[\Psi^{\dagger}\Psi+\Phi^{\dagger}\Phi]\right),
\label{eq:Z_CRMT}
\end{eqnarray}
where
\begin{equation}
D(\mu)+m=\left(\begin{array}{cc}
m & i\cosh(\mu)\Phi+\sinh(\mu)\Psi\\
i\cosh(\mu)\Phi^{\dagger}+\sinh(\mu)\Psi^{\dagger} & m
\end{array}\right).
\label{eq:D_CRMT}
\end{equation}
The degrees of freedom of the model are $N \times N$ general 
complex matrices $\Psi$ and $\Phi$.
Since its introduction it has attracted
attention due to the many features which it shares with QCD
\cite{CRMT1,CRMT2,CRMT3}: they both
have in their functional integral the determinant of a Dirac operator 
and the flavor symmetries and explicit breaking hereof are identical. 
Chiral perturbation theory at leading order in the $\epsilon$-domain 
is the relevant low energy theory in the microscopic limit for both 
theories, which resulted in a lot of interesting insights into QCD 
coming from the (much simpler) random matrix theory. The microscopic 
limit in which contact is made with $\epsilon$-regime of chiral 
perturbation theory is that of $N \rightarrow \infty$ with 
$\tilde{m} \equiv Nm$ and $\tilde{\mu} \equiv \sqrt{N}\mu$ kept
constant. \\
A sign problem is there for this theory as it is for 
QCD. This sign problem can be a severe one, as it is made manifest 
by considering the observable we will 
be concerned with, \ie the mass dependent chiral condensate
\begin{equation}
\frac{1}{N}\left\langle \bar{\eta}\eta\right\rangle =\frac{1}{N}\partial_{m}\log\left(Z\right).
\label{eq:condensate}
\end{equation}
Figure \ref{fig:PhaseQuenchedFailure} displays both the exact 
(solid red line) and the phase quenched (dashed blue line) results for 
$\frac{1}{N}\left< \bar{\eta}\,\eta\right>$ as a function of $\tilde{m}$ 
for $N=1,2,3,4$ and fixed $N_f=2$, $\tilde{\mu}=2$. 
As it can be seen, the sign problem is 
indeed severe in certain regimes of small (rescaled) 
masses\footnote{The value of the condensate is real. This 
has to be understood later when we will compare to our results; we
will always plot only the real part, the imaginary one having been 
correctly verified to be zero within errors.}. \\

Our interest in the model was triggered by \cite{KimRMT1,KimRMT2}: the
nature of the sign problem (which is due to the determinant) has a 
counterpart in a non-trivial success of the complex Langevin method 
(which needs to take the logarithm of the determinant to define the
standard effective action dictating the drift term of Langevin equation). 
While the application of complex Langevin in the most 
direct parametrization of the theory fails \cite{KimRMT1}, a different
parametrization (resulting in a different complexification) reproduces
the right results \cite{KimRMT2}.

\subsection{How many thimbles should we take into account?}

We take the most direct path to complexification, \ie for each 
field (we directly deal with the matrix elements) 
$\Phi_{ij} = a_{ij} + i b_{ij}$ and 
$\Psi_{ij} = \alpha_{ij} + i \beta_{ij}$, each real component 
gets complexified ({\em e.g.} 
$\beta_{ij} = \beta_{ij}^{(R)} + i \beta_{ij}^{(I)}$). We adhere to 
the notation of \cite{KimRMT1} and denote the action as
\begin{equation*}
S\left(a,b,\alpha,\beta\right)=N\sum_{i,j}\left(a_{ij}^2+b_{ij}^2+\alpha_{ij}^2+\beta_{ij}^2\right)-
N_f\mrm{Tr}\log\left(m^2\mathbb{1}_{N\times N}-XY\right)
\end{equation*}
with
\begin{eqnarray*}
X_{ij} &=& i\cosh\mu\left(a_{ij}+ib_{ij}\right)+\sinh\mu\left(\alpha_{ij}+i\beta_{ij}\right)\\
Y_{ij} &=& i\cosh\mu\left(a_{ji}-ib_{ji}\right)+\sinh\mu\left(\alpha_{ji}-i\beta_{ji}\right)
\end{eqnarray*}
Once we have complexified the degrees of freedom, the first step for 
the thimble approach is the identification of critical points 
of the resulting action. 
First candidate is the absolute minimum which is
already there for the real formulation, \ie $\Psi = \Phi = 0$.
All the relevant formulae for the spectral analysis of the hessian 
of $S_R$ are collected in appendix \ref{sec:RMThessian}. 
Here we 
simply state that the hessian in $0$ has the expected number of
positive eigenvalues, \ie the real
dimension of the thimble attached to $0$ is $4N^2$. Note that there is
a huge degeneracy: we have only two different eigenvalues, with the
two eigenspaces having the same dimension. As the (rescaled) mass
$\tilde{m}$ gets smaller, the gap between the two eigenvalues gets
larger. Some insight can now be gained from equation 
(\ref{eq:asymptSOL}): in first approximation, the closer the
eigenvalues, the more isotropic we expect the thimble to be. This
expectation turned out to be correct in view of the results of 
our simulations. \\

We tried to identify other critical points. In our study we explored
different values of $\tilde{m}$ (at different values of $N$) while
keeping fixed $N_f=2$ and $\tilde{\mu}=2$. One approach was solving
$\nabla S=0$ via Newton-Raphson method. We cross-checked results 
by applying the Nelder-Mead simplex method to minimize 
$\parallel \nabla S \parallel^2$. We found two classes of extrema, 
both outside the original domain and featuring an action smaller than 
$S_R(\Psi = \Phi = 0)$, which turns out to be the absolute
minimum in the original domain. Under such conditions, since the
unstable thimbles attached to the extrema we found can not intersect
the original domain of integration, we expect no contribution from
their stable thimbles ($n_\sigma=0$; see section II.B.3 of 
\cite{OurFirstTHMBL} for a more extensive discussion).

\subsection{Algorithmic issues for the CRM model}

While for the $0$-dim toy model the original algorithmic solution 
proposed in \cite{OurFirstTHMBL} is trivial, this is not the case for
the CRM model. However, previous experience with the Bose gas 
\cite{BoseTaming} taught us that there can be lucky cases. 
Let us remind the reader of the Aurora algorithm and of its gaussian
approximation (which successfully deals with the lucky cases we were
referring to).\\
We want to extract a proper noise vector for the Langevin dynamics
$$
\frac{\mrm{d}\phi_i}{\mrm{d}t}=- \frac{\de S_R}{\de\phi_i} + \eta_i \qquad i=1\cdots 2n
$$
We can proceed as follows \cite{OurFirstTHMBL}:
\begin{itemize}
\item We extract a gaussian noise $\eta_i^{(0)}(0)$, where the 
superscript qualifies this quantity as an initial proposal and the
argument has to be thought as a flow time in a sense that will be
clear soon.
\item We evolve it following the flow (\ref{eq:VecTransport})
  downwards (\ie with a change of sign with respect to 
(\ref{eq:VecTransport})), aiming at getting close enough to the 
critical point in order to make contact with the regime of 
(\ref{eq:asymptSOL}). This will hold at a given descent time 
$\tau^*$.
\item We then project with 
\beq
\eta_i^\parallel = P_{ij} \; \eta_j^{(0)}(\tau^*) \qquad P \equiv \frac{1}{2}\left(\frac{H}{\sqrt{H^2}} + 1\right)
\eeq
and normalize the result
\beq
\eta(\tau^*) = r \; \frac{\eta^\parallel}{\|\eta^\parallel\|}
\eeq
r being extracted according to the n-dimensional $\chi$ distribution.
\item We then ascend along the flow, covering again a time interval of 
length $\tau^*$. The result is the noise $\eta_i$ we will put in our
Langevin equation.
\end{itemize}
Extracting the noise vector is not yet the end of the story, since any
finite order approximation to Langevin equation, e.g. the Euler scheme 
$$
\phi_i' = \phi_i - \delta t \; \frac{\de S_R}{\de\phi_i} +
\sqrt{\delta t} \; \eta_i
$$
will introduce systematic effects; since the manifold is not flat, the
final point $\phi_i'$ will be moved away from the thimble. The obvious
remedy for this effect is to repeat just the same we did for the noise
vector (move the configuration along the flow downward, close to the
critical point, project it onto the tangent space, move it upward
along the flow for the same time length). Note that it is expected
that, in all the descent/ascent mechanisms we have just described, 
the downward flow, \ie the SD will be numerically delicate. It is thus 
much better to formulate the descent as a boundary value problem 
(BVP) rather than as an initial value problem, as it was observed 
in \cite{BoseTaming}.\\

A much more appealing observation was also made in \cite{BoseTaming}: 
there are lucky cases in which a quite rough approximation holds; 
with a slight linguistic abuse we call it a {\em gaussian 
approximation}\footnote{One should not confuse this with the gaussian
approximation described in 3.1. In that case the action is
approximated with its leading (gaussian) term and all the thimble
analysis is performed consequently. In this (algorithmic) gaussian
approximation one pretends that the thimble manifold we are interested
in and the thimble associated to the gaussian approximation of the
action sit on top of each other also away from the critical point.}.
Roughly speaking, this means taking the minimum value for the 
$\tau^*$ technical parameter, \ie $\tau^*=0$. This formally relies 
on the assumption that integrating the system on the vector space
defined by the tangent space at the critical point actually takes into
account the relevant configurations giving the most important
contribution to the functional integral. This was actually
holding in the case of \cite{BoseTaming}. \\

\begin{figure}[ht]
\begin{center}
\includegraphics[height=9cm,clip=true]{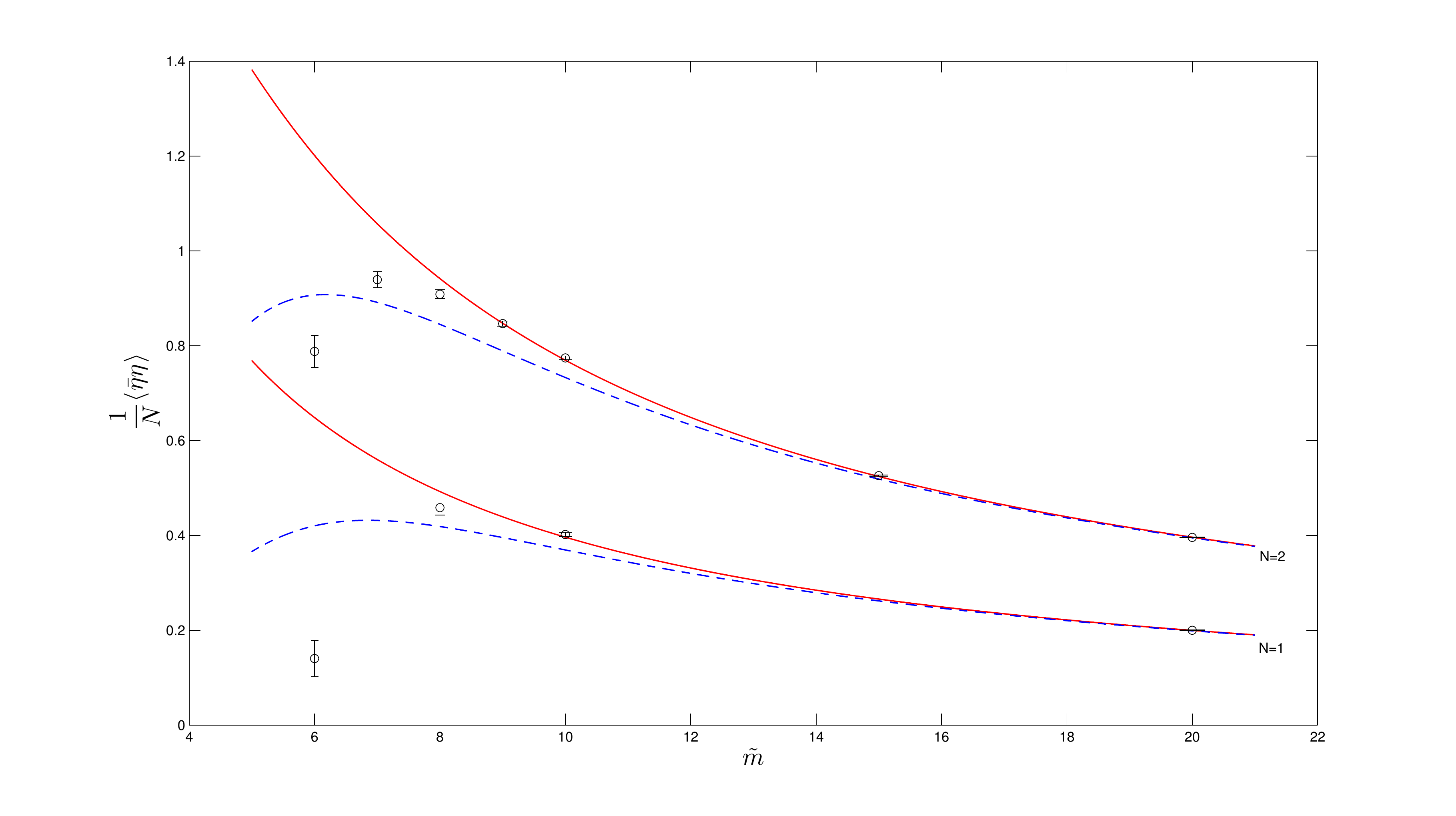}
\caption{Exact (solid red line), phase quenched 
(dashed blue line) and gaussian approximation 
results for the condensate, 
at fixed $N_f=2$, $\tilde{\mu}=2$.}
\label{fig:CRMgaussapprx}
\end{center}
\end{figure}

Does the gaussian approximation hold true also for the CRM model?
Figure \ref{fig:CRMgaussapprx} reveals that there is actually a regime in
which it can do pretty well. Not surprisingly, it is a regime in which
results are not that far away from the phase quenched approximation;
we know that this is a regime in which the two different eigenvalues 
of the hessian at the critical point are quite close to each other and 
the problem appears all in all quite symmetric and not that far away 
from the regime of (\ref{eq:asymptSOL}). Note that the value of the
(rescaled) mass at which the solution provided by the gaussian 
approximation departs from the correct one varies with $N$. \\

Next step was to leave the gaussian approximation aside and try to
implement the full Aurora algorithm. There are a couple of issues one
should be aware of: we need a solid estimate for $\tau^*$; also, within a
time length of order $\tau^*$ we have to make sure we have under good
numerical control both the SA and the SD. The latter is the
critical one, for which we have already made clear that a BVP
formulation is the choice to go for. Our implementation was along the 
same lines of the code available at \cite{thimbleCODE}. All in all,
our experience with the complete Aurora algorithm for the CRM model
was at first somehow inconclusive: a clear-cut indication of values of
$\tau^*$ at the same time safe and manageable was missing. We
will come back to this observation later, in the framework of the
other numerical approach that we chose to implement. 

\subsection{A different numerical approach}

We now want to take advantage of the parametrization
$$
\Phi\in\mcal{J}_\sigma\leftrightarrow\left(\hat{n},t\right)
$$
(a few of the formulae we will need to implement our strategy 
were clearly stated in \cite{Kikukawa}, while the strategy itself 
we will see at work in the following was first described in 
\cite{Lat2014algo}). Basically one describes a generic point by
locating it on the SA curve it lies on. This means 
providing $\hat{n}$ (the direction one is taking while leaving the
critical point) and the time $t$ at which one reaches $\Phi$ 
while integrating the SA equations. The first goal is now
to rewrite the contribution to the partition function which is
attached to one thimble. In full detail this reads 
\beqn 
\nonumber Z ^{\left(\sigma\right)} & = &\int_{\mathcal{J}_{\sigma}} dz_1 \wedge \ldots \wedge dz_n \; \mbox{e}^{-S} \\
& = & \sum_{\mbox{\tiny{charts}} \,c} \, \int_{\Gamma_c} \prod_i^n dy_i^c \; \mbox{det}(U) \; \mbox{e}^{-S} = 
\mbox{e}^{-i S_I} \sum_{\mbox{\tiny{charts}} \,c} \, \int_{\Gamma_c} \prod_i^n dy_i^c \; \mbox{e}^{i\omega} \; \mbox{e}^{-S_R}
\label{eq:fullZ}
\eeqn
In (\ref{eq:fullZ}) we have taken into account that $S_I$ is constant
on $\mathcal{J}_{\sigma}$. Moreover, there could be
more than one relevant chart and on a given chart we have to take
into account the residual phase. For the sake of simplicity in
notation, we now take a few shortcuts. First of all, we discard 
the overall phase $\mbox{e}^{-i S_I}$; it will be easy to account for
it when we come back to the actual computation of an observable. 
We also discard the fine detail of more than one chart, since in
practice this is not a issue\footnote{We will be concerned with single
ascents, for which we will revert to a different, smooth parametrization.}. 
Finally, we leave the
residual phase aside, having in mind that we can take it into account
{\em a posteriori} by reweighting. We thus write a new quantity, which
is the one we will further manipulate in order to single out the
contributions from the single ascents. We define\footnote{For the sake
of notational simplicity we also omit the explicit indication that the
integration is on the thimble, as it is easy to recognize we are
assuming. Notice that (\ref{eq:Zsgm}) is the generalization of 
(\ref{eq:Zs0}) of section 3.1.}
\beq
\mcal{Z}^{\left(\sigma\right)} \equiv \int\prod_{i=1}^n\mrm{d}y_i\,e^{-S_R}
\label{eq:Zsgm}
\eeq
Roughly speaking, this is the quantity that can have a probabilistic
interpretation. The key point is now to write an expression for $1$
\beq
1=\Delta_{\hat{n}}\left(t\right)\int\prod_{k=1}^n\mrm{d}n_k\,\delta\left(\left|\vec{n}\right|^2-1\right)\int\mrm{d}t\,\prod_{i=1}^n\delta\left(y_i-y_i\left(\hat{n},t\right)\right)
\label{eq:similFP}
\eeq
where $\{y_i\left(\hat{n},t\right)\}$ are the coordinates of the field 
as expressed in the local (orthonormal) basis 
$\{U^{\left(i\right)}_{\hat{n}}\left(t\right)\}$ 
parallel-transported along the SA defined by 
$\hat{n}$ until time $t$. The solution for $\Delta_{\hat{n}}\left(t\right)$ 
is in terms of (the module of) a determinant
\beq
\nonumber \Delta_{\hat{n}}\left(t\right)= \left| \det 
\bmat
\frac{\delta\left(\left|\vec{n}'\right|^2-1\right)}{\delta t} & \frac{\delta\left(\left|\vec{n}'\right|^2-1\right)}{\delta n_1} & \cdots & \frac{\delta\left(\left|\vec{n}'\right|^2-1\right)}{\delta n_n}\\
\frac{\delta\left(y_1-y_1\left(\hat{n},t\right)\right)}{\delta t} & \frac{\delta\left(y_1-y_1\left(\hat{n},t\right)\right)}{\delta n_1} & \cdots & \frac{\delta\left(y_1-y_1\left(\hat{n},t\right)\right)}{\delta n_n}\\
\vdots & \vdots & \ddots & \vdots\\
\frac{\delta\left(y_n-y_n\left(\hat{n},t\right)\right)}{\delta t} & \frac{\delta\left(y_n-y_n\left(\hat{n},t\right)\right)}{\delta n_1} & \cdots & \frac{\delta\left(y_n-y_n\left(\hat{n},t\right)\right)}{\delta n_n}\\
\emat
\right|
\eeq
or
\beq
\Delta_{\hat{n}}\left(t\right)= \left| \det 
\bmat
0 & 2 n_1 & \cdots & 2 n_n\\
\frac{\delta y_1\left(\hat{n},t\right)}{\delta t} & \frac{\delta y_1\left(\hat{n},t\right)}{\delta n_1} & \cdots & \frac{\delta y_1\left(\hat{n},t\right)}{\delta n_n}\\
\vdots & \vdots & \ddots & \vdots\\
\frac{\delta y_n\left(\hat{n},t\right)}{\delta t} & \frac{\delta y_n\left(\hat{n},t\right)}{\delta n_1} & \cdots & \frac{\delta y_n\left(\hat{n},t\right)}{\delta n_n}\\
\emat
\right|
\eeq

The first column of this determinant can be easily related to 
the gradient of the action. It turns out that to compute the 
generic matrix element we need to do the following: \\

\begin{itemize}
\item We need to evolve not only the field, but the entire basis 
by integrating (\ref{eq:VecTransport}). 
\item We construct the $2n\times n$ matrix $V$ whose columns are 
the $\{V^{\left(i\right)}\left(t\right)\}$. 
\item We construct the $2n\times n$ matrix $u$ whose columns are
the vectors $\{u^{\left(i\right)}\}_{i=1\cdots n}$ which are obtained from the 
$\{V^{\left(i\right)}\left(t\right)\}$ by means of Gram-Schmidt orthonormalization 
procedure. 
\item The relation $V=uE$ holds, with 
$$
E_{ij}=
\bcas
V^{\left(j\right)}\cdot u^{\left(i\right)} & j\geq i\\
0 & j<i
\ecas
$$
\item The entries of the determinant we are looking for are now given 
by
\beq
\bcas
\frac{\delta y_i}{\delta t} = \sum_{k=1}^n\lambda_k n_k E_{ik}\\
\frac{\delta y_i}{\delta n_j} = E_{ij}
\ecas
\label{eq:DLTentries}
\eeq
\end{itemize}
Not surprisingly, there is a lot of information in the 
(tremendous amount of) computations we have just sketched. 
In particular, if we now introduce the $n\times 2n$ 
complex space projector $P$
\beq
P=
\bmat
\mbb{1}_{n\times n}&i\,\mbb{1}_{n\times n}
\emat
\eeq
then the $n\times n$ complex matrix $U=Pu$ is unitary; 
this is precisely the matrix of eq.~(\ref{eq:RESphase}), 
whose determinant is the residual phase $e^{i\omega}$. \\

The details of the previous computation of $\Delta_{\hat{n}}\left(t\right)$
are given in appendix \ref{sec:DeltaCOMP}. We now proceed to make use of the 
expression for the identity encoded in (\ref{eq:similFP}).
Inserting it in (\ref{eq:Zsgm}) we get
\beqn
\nonumber\mcal{Z}^{\left(\sigma\right)}&=&\int\prod_{i=1}^n\mrm{d}y_i\,e^{-S_R}\\
\nonumber&=&\int\prod_{i=1}^n\mrm{d}y_i\,e^{-S_R}\Delta_{\hat{n}}\left(t\right)\int\prod_{k=1}^n\mrm{d}n_k\,\delta\left(\left|\vec{n}\right|^2-1\right)\int\mrm{d}t\,\prod_{i=1}^n\delta\left(y_i-y_i\left(\hat{n},t\right)\right)\\
\nonumber&=&\int\prod_{k=1}^n\mrm{d}n_k\,\delta\left(\left|\vec{n}\right|^2-1\right)\int\mrm{d}t\int\prod_{i=1}^n\mrm{d}y_i\,\delta\left(y_i-y_i\left(\hat{n},t\right)\right)\Delta_{\hat{n}}\left(t\right) e^{-S_R}\\
\nonumber&=&\int\prod_{k=1}^n\mrm{d}n_k\,\delta\left(\left|\vec{n}\right|^2-1\right)\int\mrm{d}t\,\Delta_{\hat{n}}\left(t\right) e^{-S_R\left(\hat{n},t\right)}
\eeqn
which has a possible interpretation in terms of
\beq
\mcal{Z}^{\left(\sigma\right)}=\int\prod_{k=1}^n\mrm{d}n_k\,\delta\left(\left|\vec{n}\right|^2-1\right)\mcal{Z}^{\left(\sigma\right)}_{\hat{n}}
\eeq
\ie there is a contribution to the partition function for each
SA path
\beq
\mcal{Z}^{\left(\sigma\right)}_{\hat{n}}=\int\limits_{-\infty}^{+\infty}\mrm{d}t\,\Delta_{\hat{n}}\left(t\right) e^{-S_R\left(\hat{n},t\right)}
\eeq
Note that the procedure  naturally defines a probability, \ie that
for a point reached at time $t$ on the SA defined by $\hat{n}$ 
\beq
P_{\hat{n}}\left(t\right)=\frac{\Delta_{\hat{n}}\left(t\right) e^{-S_R\left(\hat{n},t\right)}}{\mcal{Z}^{\left(\sigma\right)}_{\hat{n}}}
\eeq
One can also naturally define the cumulative distribution function
(it is manifestly non-decreasing, positive definite and has the 
correct normalization)
\beq
F_{\hat{n}}\left(t\right)=\frac{1}{\mcal{Z}^{\left(\sigma\right)}_{\hat{n}}}\int\limits_{-\infty}^t\mrm{d}t'\,\Delta_{\hat{n}}\left(t'\right) e^{-S_R\left(\hat{n},t'\right)}
\eeq
Since we can easily invert this function numerically, we have a tool 
to ideally sample configurations on a single SA. Namely, 
we extract a random number $\xi\in\left[0,1\right]$ and then get the point on 
the SA (rather, the time at which the point is reached) 
by $t=F_{\hat{n}}^{-1}\left(\xi\right)$. Actually this is not that useful.
The fact that we ascend all the way along a given SA in order to
compute $\mcal{Z}^{\left(\sigma\right)}_{\hat{n}}$ suggests that it is rather
convenient to compute the entire contribution 
which is attached to that given ascent. On the other hand, 
the relative weight of a given SA (within the complete 
partition function $\mcal{Z}^{\left(\sigma\right)}$) is given by 
$\mcal{Z}^{\left(\sigma\right)}_{\hat{n}}/\mcal{Z}^{\left(\sigma\right)}$. \\

We now want to take advantage of the parametrization 
$\Phi\in\mcal{J}_\sigma\leftrightarrow\left(\hat{n},t\right)$ in the
computation of an observable. In the following, we will assume 
we are in a case in which only one single thimble is relevant. 
This is not the general case, but for what we want to obtain it is not
a limitation. In the cases in which more than one thimble contribute,
we can address the problem using the same strategy described in
3.1 in the context of the quartic toy model: it will be easy for the
reader to generalize eq.~(\ref{eq:MANYthimbles}) and the 
discussion following it. 
With this caveat in mind, we first of all write 
\beqn
\nonumber \left< O \right> & = & 
\frac{\int_{\mathcal{J}_{\sigma}} dz_1 \wedge \ldots \wedge dz_n \; O  \;\mbox{e}^{-S}}{Z ^{\left(\sigma\right)}} \\
\nonumber & = & \frac{\int\prod_{i=1}^n\mrm{d}y_i\; O \; e^{i \omega} \,e^{-S_R}}
{\int\prod_{i=1}^n\mrm{d}y_i \; e^{i \omega} \,e^{-S_R}} =
\frac{\langle e^{i \omega} \; O\rangle_\sigma}
{\langle e^{i \omega}\rangle_\sigma}
\eeqn
where $\langle \dots \rangle_\sigma \, \equiv \, \mcal{Z}^{\left(\sigma\right) -1}
\int\prod_{i=1}^n\mrm{d}y_i\, \dots \,e^{-S_R} $. We have till now
simply generalized eq.~(\ref{eq:VEVs0}). We can go
further by making use of the new parametrization we 
introduced\footnote{We denote ${\cal{D}}\hat{n} \equiv \prod_{k=1}^n\mrm{d}n_k\,\delta\left(\left|\vec{n}\right|^2-1\right)$.}
\beqn
\nonumber <O> & = & \frac{\int {\cal{D}}\hat{n} \int dt \;
  \Delta_{\hat{n}}(t) \,e^{-S_R(\hat{n},t)} \; e^{i \omega(\hat{n},t)}
  \; O(\hat{n},t) }{\int {\cal{D}}\hat{n} \int dt \;
  \Delta_{\hat{n}}(t) \; e^{-S_R(\hat{n},t)} \, e^{i \omega(\hat{n},t)}} \\
\nonumber & = & \frac{\int {\cal{D}}\hat{n} \; {\cal{Z}}^{(\sigma)}_{\hat{n}}\left( {\cal{Z}}^{(\sigma)-1}_{\hat{n}} \; \
\int dt \; \Delta_{\hat{n}}(t) \; \mbox{e}^{-S_R(\hat{n},t)}  \; e^{i
  \omega(\hat{n},t)}\; O(\hat{n},t) \right)}
{\int {\cal{D}}\hat{n} \; {\cal{Z}}^{(\sigma)}_{\hat{n}} \left( {\cal{Z}}^{(\sigma)-1}_{\hat{n}} \; \
\int dt \; \Delta_{\hat{n}}(t) \; \mbox{e}^{-S_R(\hat{n},t)}  \; e^{i
  \omega(\hat{n},t)} \right)} \\
& \equiv & \frac{\int {\cal{D}}\hat{n} \;
  {\cal{Z}}^{(\sigma)}_{\hat{n}}\left< e^{i \omega} \, O \right>_{\hat{n}}}{\int {\cal{D}}\hat{n} \; {\
\cal{Z}}^{(\sigma)}_{\hat{n}} \left< e^{i \omega} \right>_{\hat{n}} }
\label{eq:newINT}
\eeqn
Eq. (\ref{eq:newINT}) is in a sense a new average. Namely, the different 
directions $\hat{n}$ can now be regarded as the new degrees 
of freedom of the overall integral, the quantities to be measured 
are the $\left< O \,e^{i \omega}\right>_{\hat{n}}$ and 
$\left< e^{i \omega} \right>_{\hat{n}}$ (\ie partial averages 
attached to single SA), and the weights are given by the 
$\mcal{Z}^{\left(\sigma\right)}_{\hat{n}}$. 
It is rather obvious that
\begin{itemize}
\item The basic building block are complete ascents. This is 
good, since we can have their computation under good numerical 
control. In other words, sampling {\em on} the thimble is not a 
problem: we stay on the thimble by definition.
\item The way to importance sampling now appears tricky. This 
is easy to understand, since picking up a contribution 
means picking up a $\hat n$, whose weight 
$\mcal{Z}^{\left(\sigma\right)}_{\hat{n}}$ is not known {\em a
  priori}, but only after the SA 
path associated to $\hat n$ has been obtained.
\item The crudest approach one can think of is of course a 
uniform sampling of the $\hat{n}$-space; this is a static, 
crude Monte Carlo, which can easily become inefficient (in 
particular for large systems).
\end{itemize}
In the following we will just be satisfied with the last 
approach of static, crude Monte Carlo: this will be enough to 
show that we
can reproduce the correct results for the model at hand 
(even in regions where the sign problem is quite severe) and this 
holds true taking into account the contribution of the single 
critical point we found. From this very basic approach there will 
be something to be learnt also with respect to the 
Aurora algorithm (and on the computation of density of
states as well). We will finally report on a few speculations on
smarter algorithms which we are trying to devise.

\subsection{Results for the CRM model}

\begin{figure}[ht]
\begin{center}
\includegraphics[height=9cm,clip=true]{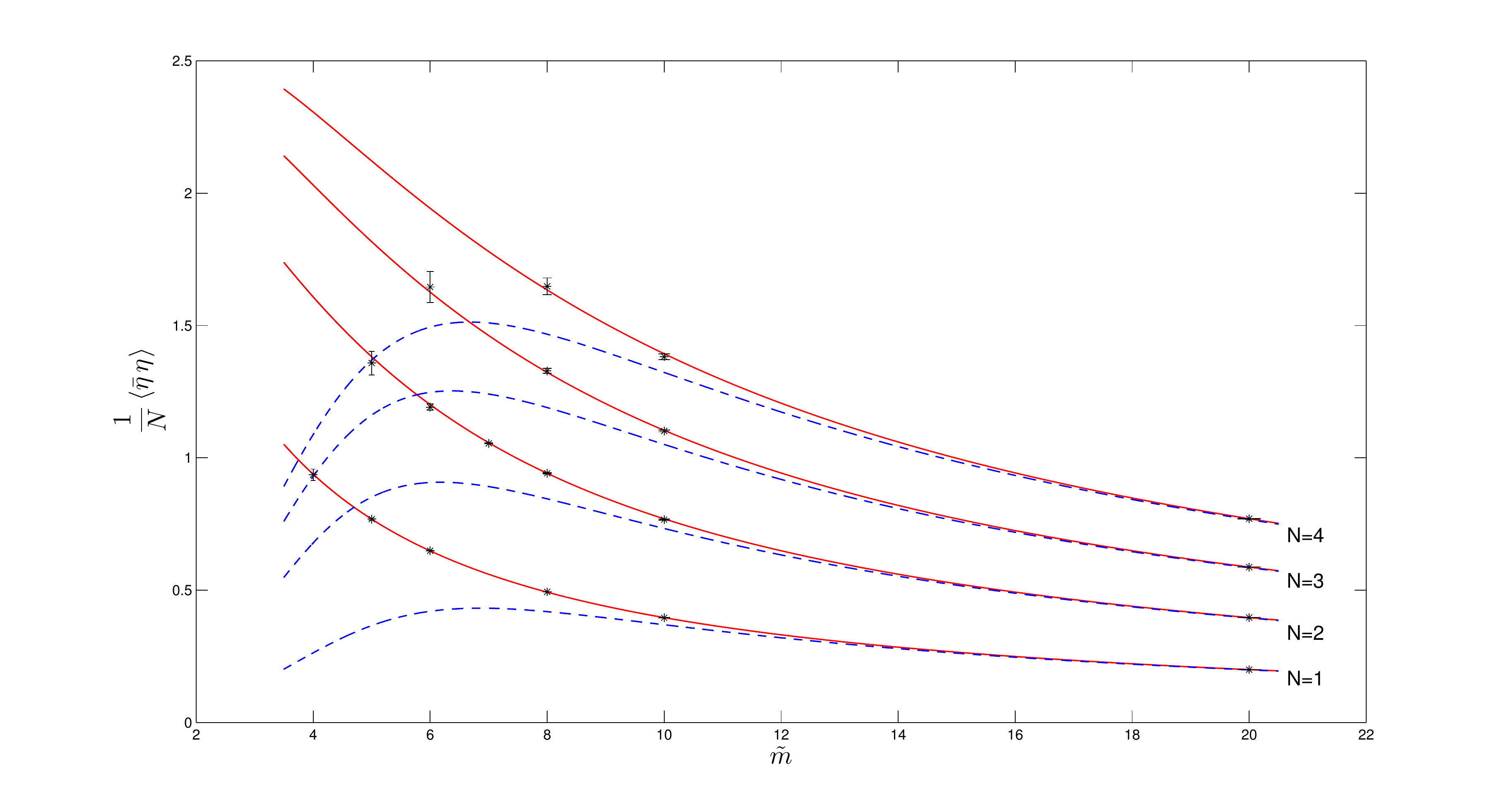}
\caption{Exact (solid red line), phase quenched 
(dashed blue line) and thimble simulations results for the condensate, 
at fixed $N_f=2$, $\tilde{\mu}=2$.}
\label{fig:finalplot}
\end{center}
\end{figure}

Figure \ref{fig:finalplot} displays the results we obtained from
simulations performed in the static Monte Carlo approach we have just 
discussed. All these results come from the contribution of one single
thimble. As we have already pointed out, one original motivation of 
ours turned out to be not relevant: we were ready for looking 
for dominance of one thimble in some asymptotic regime (in the
thermodynamic limit) and, in the region of parameters we studied, 
we actually found no other thimble but the 
trivial one. Also, complexifying the theory in the parametrization 
that was shown to be problematic for complex Langevin \cite{KimRMT1}
did not result in any problem. Results are shown
for $N=1,2,3,4$ and fixed $N_f=2$, $\tilde{\mu}=2$. \\

\begin{figure}[ht]
\begin{center}
\includegraphics[height=9cm,clip=true]{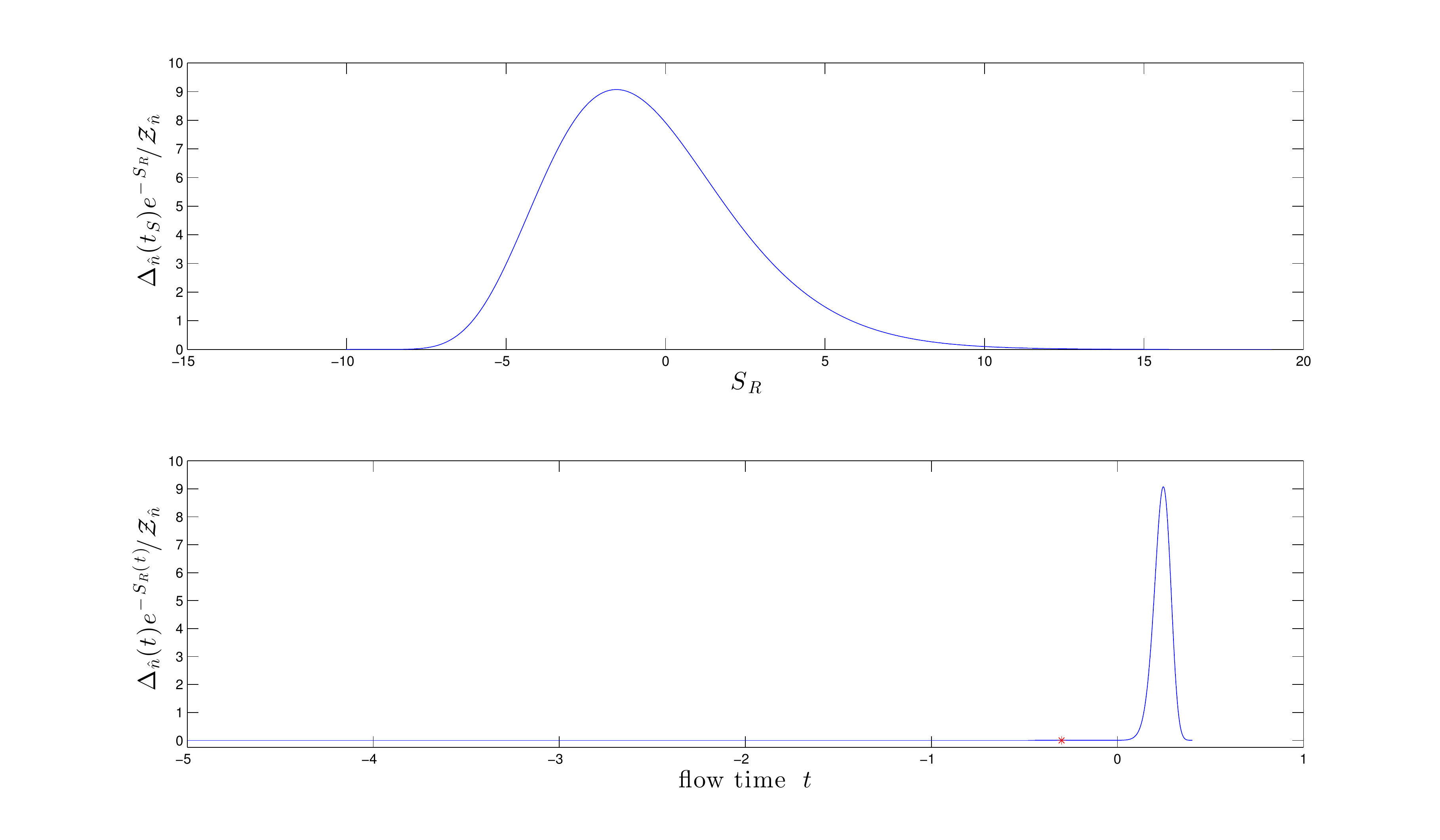}
\caption{Real weight of the 
functional integral for the configurations which lie on the SA 
defined by a particular $\hat{n}$ as a function of $S_R$ (upper panel) and
$t$ (lower panel) for $N=2$, $\tilde{m}=7$, $N_f=2$, $\tilde{\mu}=2$.}
\label{fig:N2m7slow}
\end{center}
\end{figure}

It is interesting to regard the parametrization we have employed from
another point of view. In the upper panel of 
figure~\ref{fig:N2m7slow} we plot the quantity 
$\Delta_{\hat{n}}(t) \;
\mbox{e}^{-S_R(\hat{n},t)}/\mcal{Z}^{\left(\sigma\right)}_{\hat{n}}$ 
as a function of
$S_R$ along a given ascent (remember that on each ascent one single
value of $S_R$ is only read once). This is the real weight of the 
functional integral for the configurations which lie on that given 
ascent: it can be thought of as a different way of looking at the 
density of states (namely, this is the contribution attached to a given 
ascent). \\
In the lower panel of figure~\ref{fig:N2m7slow} we plot the same
quantity as a function of the flow time. A first remark is due for 
the long initial flat region. Consider eq.~(\ref{eq:asymptSOL}). When
we want to compute the contribution from a single ascent (that is from a
single $\hat{n}$) we need an initial condition, \ie an initial value 
$t_0$ at which the asymptotic regime holds. In principle, the more 
back in time we take $t_0$, the better initial condition we
prepare. There are of course accuracy issues one has to live with. 
The flat initial region
reflects the fact that we do our best to ensure we stay on the
thimble. On the other side, one would like to know till what value of
the flow time
the asymptotic regime holds to a reasonable confidence. We can think
of more than one indicator for the latter condition, e.g. the gaussian
approximation of the action is very close to the actual value of $S$,
or the factor $\Delta_{\hat{n}}(t)$ is very close to its gaussian
approximation (see appendix \ref{sec:DeltaCOMPgss}). We mark with a (red) star a value of
flow time which can be assumed to be the boundary of the region we have
just described. Now, an efficient dynamic Monte Carlo is supposed to
sample configurations in regions where the weight is concentrated. In
this sense we can say that the distance (in flow time) from the region
around the maximum of 
$\Delta_{\hat{n}}(t) \; \mbox{e}^{-S_R(\hat{n},t)}/\mcal{Z}^{\left(\sigma\right)}_{\hat{n}}$ 
and the flow time marked with the star is a reasonable indicator of the
$\tau^*$ parameter of the Aurora algorithm. \\

\begin{figure}[hb]
\begin{center}
  \begin{tabular}{c} 
 \hspace{-0.2cm}
     \includegraphics[height=7.5cm,clip=true]{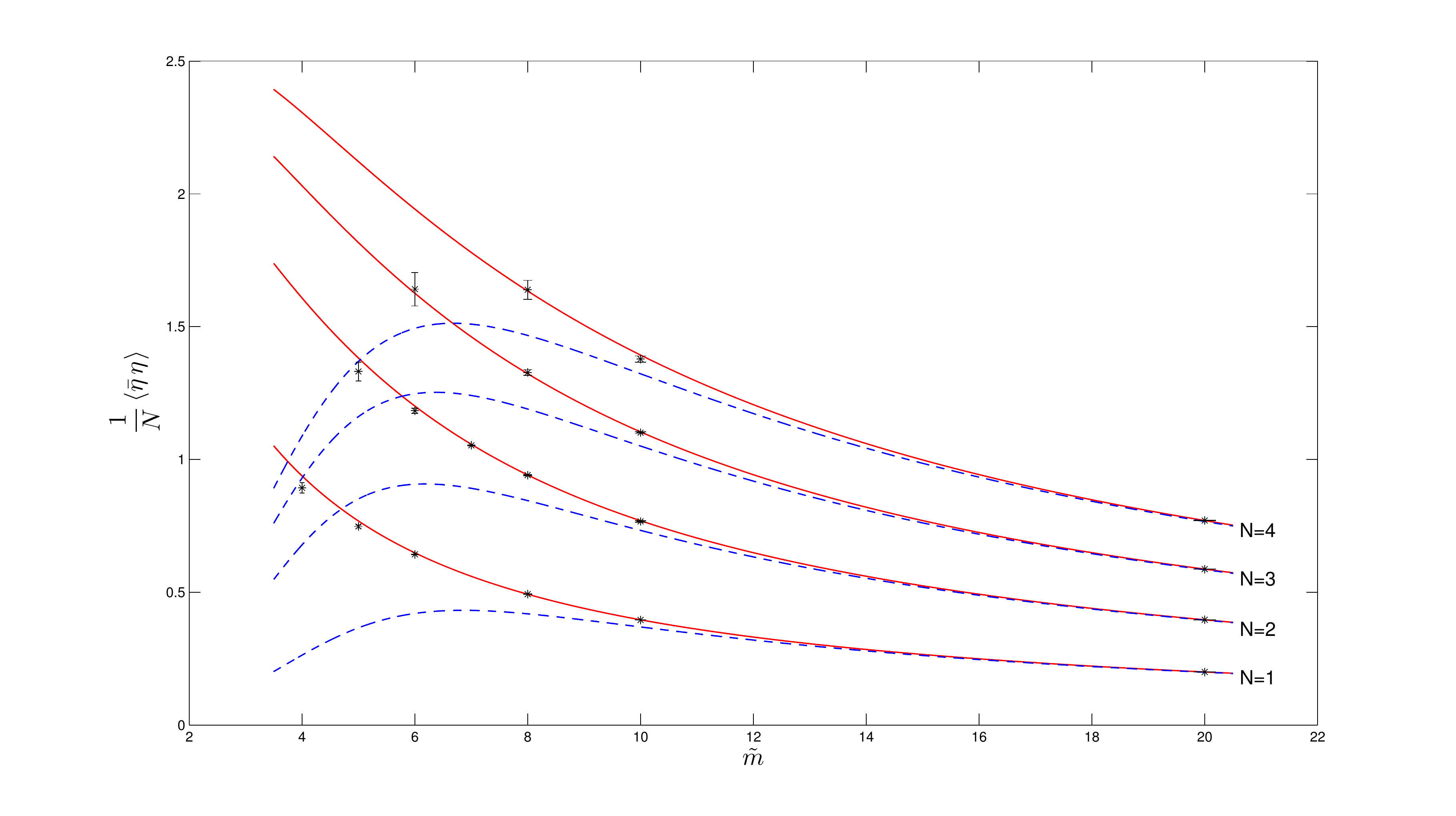} \\
 \hspace{-0.2cm}
 \vspace{-0.6cm}
     \includegraphics[height=7.5cm,clip=true]{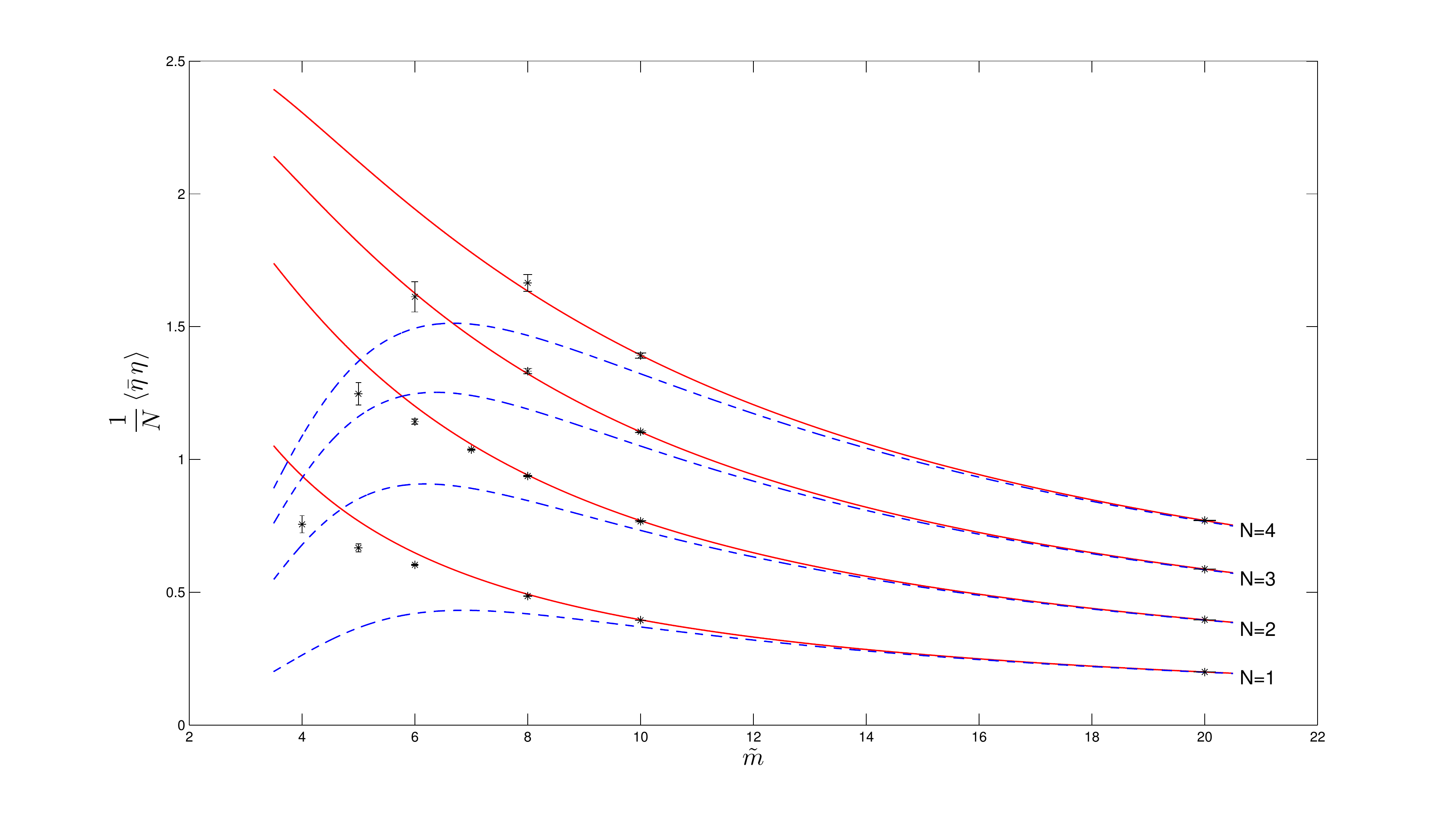}
  \end{tabular}
\end{center}
  \caption{The effect of not accounting for the residual phase. In the upper
    panel its contribution is not accounted in the denominator of 
eq.~(\ref{eq:newINT}), \ie we compute ${\cal{Z}}^{(\sigma)-1} \, \int {\cal{D}}\hat{n} \;
{\cal{Z}}^{(\sigma)}_{\hat{n}} \left< e^{i \omega} \, O
  \right>_{\hat{n}}$. In the lower panel we omit the residual phase
  completely and compute ${\cal{Z}}^{(\sigma)-1} \, \int {\cal{D}}\hat{n} \;
{\cal{Z}}^{(\sigma)}_{\hat{n}} \left< O \right>_{\hat{n}}$.}
  \label{fig:RPeffects}
\end{figure}

A comment is due concerning the residual phase: we encountered no
problem in taking it into account by reweighting. As it was expected,
it is a smooth function on the ascents, so that 
$\left< O \,e^{i \omega}\right>_{\hat{n}}$ and 
$\left< e^{i \omega} \right>_{\hat{n}}$ can be safely computed and
wild cancellations are never there at any stage of our computations. 
The fact that the residual phase is smooth does not of course mean it
has no net effect, as can be seen in figure~\ref{fig:RPeffects}. In
the upper panel we show the effect of neglecting the contribution of
the denominator of eq. (\ref{eq:newINT}): this amounts to computing 
${\cal{Z}}^{(\sigma)-1} \, \int {\cal{D}}\hat{n} \;
{\cal{Z}}^{(\sigma)}_{\hat{n}} \left< e^{i \omega} \, O
  \right>_{\hat{n}}$\footnote{Note that in this
  case we take the residual phase into
  account in the numerator.}. In the lower panel we show yet
another phase quenched computation, namely what we could term a 
{\em residual phase quenched} result. In this case we simply omit the
contribution of the residual phase and compute 
${\cal{Z}}^{(\sigma)-1} \, \int {\cal{D}}\hat{n} \;
{\cal{Z}}^{(\sigma)}_{\hat{n}} \left< O \right>_{\hat{n}}$. 
Both plots show that reweighting for the residual phase is
essential to get the correct results; this happens to be the case in
particular for low dimensions.\\

We admittedly made use of the crudest possible application of the
parametrization contained in eq.~(\ref{eq:newINT}), \ie a static, crude 
Monte Carlo sampling of the integral. It is static, \ie it is not
based on a stochastic process. It is crude,
because it does not implement importance sampling\footnote{We recall
  that the 
  Monte Carlo methods we are mostly familiar with are dynamic Monte
  Carlo in which importance sampling is obtained via convergence to
  the equilibrium distribution of a stochastic process, which in 
virtually all the cases is a Markov chain.}. In general, importance
sampling computes an integral $I = \int dx f(x)$ as an average 
$I=\int dx \, \rho(x) \, \frac{f(x)}{\rho(x)}$ and all the point is
being able to extract configurations distributed according 
to $\rho(x)$. The natural importance
sampling for (\ref{eq:newINT}) would try to sample the space of
ascents (\ie the different $\hat{n}$) according to the weights 
$\mcal{Z}^{\left(\sigma\right)}_{\hat{n}}/\mcal{Z}^{\left(\sigma\right)}$ 
and we have already made the point that this is tricky. Crude 
Monte Carlo simply extracts $\hat{n}$ with constant probability. 
As it is well known, the more non-trivial the profile of
$\mcal{Z}^{\left(\sigma\right)}_{\hat{n}}/\mcal{Z}^{\left(\sigma\right)}$
is, the more inefficient one expects the crude Monte Carlo to be.
This can be clearly seen at low masses, where the problem is less
symmetric with respect to different choices of the $\hat{n}$: 
we have already made this point while commenting on the failure of the
gaussian approximation for the Aurora algorithm. The interested reader
is referred to appendix \ref{sec:DeltaCOMPgss} to get more insight on
this in the case of the gaussian action. Here we want to stress that
the problem with crude Monte Carlo does not necessarily relate to the
dimension $N$ (of the matrices): going low enough in mass at any fixed
$N$ can already result in a difficult computation\footnote{Of course
  going to higher values of $N$ would result in extra computational
  effort, but what is really crucial is to see at each value of $N$
  what is the threshold in mass below which we have a {\em too much
    non-trivial} profile of the 
$\mcal{Z}^{\left(\sigma\right)}_{\hat{n}}/\mcal{Z}^{\left(\sigma\right)}$.}. 
As the mass gets lower, one indeed 
clearly sees larger error bars in figure \ref{fig:finalplot}. Needless
to say, this is a region in which we had to collect many ascents:
hundreds of thousands, actually, by definition all statistically 
independent. This is a huge numerical effort. On the other side, 
even accepting that at some point our crude Monte Carlo has to give up
in front of a non-trivial profile of the weights 
$\mcal{Z}^{\left(\sigma\right)}_{\hat{n}}/\mcal{Z}^{\left(\sigma\right)}$, it is
reassuring to see that we nevertheless solved a non-trivial problem: 
figure \ref{fig:finalplot} clearly shows that we were able to solve
the problem also in regions in which the sign problem shows up as
quite severe.\\

All in all, we saw that implementing importance sampling is hard, 
since the relative weights 
$\mcal{Z}^{\left(\sigma\right)}_{\hat{n}}/\mcal{Z}^{\left(\sigma\right)}$
are only known after the complete SA associated to $\hat n$ has 
been computed. This could sound a very pessimistic conclusion. 
On the other hand, the 
gaussian approximation of the $\mcal{Z}^{\left(\sigma\right)}_{\hat{n}}$ can
be easily computed (see appendix \ref{sec:DeltaCOMPgss}), which
suggests the idea of making
use of them to formulate proposals for the $\hat{n}$. This is
something we are currently investigating. 
 
\section{Conclusions and prospects}

We discussed the solution of a simple toy model via thimble
regularization. Quite interestingly, this model, which dates 
back to some thirty years ago and was proposed as a sort of 
benchmark for complex Langevin, was still missing a full solution in
the context of the latter. In thimble regularization the solution is clear
and can be implemented numerically by a number of simulation
algorithms. \\ 
We then investigated the Chiral Random Matrix model and 
showed that thimble regularization can successfully deal with the sign
problem that the system displays. In the region of parameters we
studied, a single thimble accounts for the results. 
We made use of a parametrization in terms of contributions
attached to SA (which are the basic building blocks to
define the thimble). This was done by crude Monte Carlo, leaving 
open the problem of devising a smarter algorithm 
(importance sampling) to take advantage
of the parametrization we made use of. This is the subject we are
currently investigating in view of other applications. 

\section*{Acknowledgments}
\par\noindent
We warmly thank Luigi Scorzato for many valuable discussions and 
for all the common work on the subject in recent years.
We are indebted to Kim Splittorff for very fruitful conversations and
for having introduced us to the subject of the CRM model. 
We also acknowledge useful discussions with Marco Cristoforetti and 
Michele Brambilla and we are grateful to Christian Torrero 
who has collaborated with us at an early stage of this work.
We finally thank A. Alexandru for a useful conversation which made us 
consider a more extended comment on the effects of the residual phase. 
This research was initially supported by the Research 
Executive Agency (REA) of the European Union under Grant Agreement 
No. PITN-GA-2009-238353 (ITN STRONGnet) and by Italian 
MURST under contract PRIN2009 (20093BMNPR 004). 
We acknowledge partial support from I.N.F.N. 
under the research project {\sl i.s. QCDLAT}. 

\newpage
\appendix

\section{The Hessian for the CRM model}
\label{sec:RMThessian}







We want to compute the hessian at the critical point $a_{ij}=b_{ij}=\alpha_{ij}=\beta_{ij}=0$. We need the following second derivatives (the fields are complexified by setting \emph{e.g.} $a_{ij}=a_{ij}^R+i\,a_{ij}^I$ etc.)

\begin{eqnarray*}
\frac{\de^2S_R}{\de a_{mn}^R\de a_{ij}^R}\biggr|_0 &=& -\frac{\de^2S_R}{\de a_{mn}^I\de a_{ij}^I}\biggr|_0 = A_-\delta_{mi}\delta_{nj}\\
\frac{\de^2S_R}{\de a_{mn}^R\de a_{ij}^I}\biggr|_0 &=& \frac{\de^2S_R}{\de a_{mn}^I\de a_{ij}^R}\biggr|_0 = 0\\
\frac{\de^2S_R}{\de b_{mn}^R\de b_{ij}^R}\biggr|_0 &=& -\frac{\de^2S_R}{\de b_{mn}^I\de b_{ij}^I}\biggr|_0 = A_-\delta_{mi}\delta_{nj}\\
\frac{\de^2S_R}{\de b_{mn}^R\de b_{ij}^I}\biggr|_0 &=& \frac{\de^2S_R}{\de b_{mn}^I\de b_{ij}^R}\biggr|_0 = 0\\
\frac{\de^2S_R}{\de \alpha_{mn}^R\de \alpha_{ij}^R}\biggr|_0 &=& -\frac{\de^2S_R}{\de \alpha_{mn}^I\de \alpha_{ij}^I}\biggr|_0 = A_+\delta_{mi}\delta_{nj}\\
\frac{\de^2S_R}{\de \alpha_{mn}^R\de \alpha_{ij}^I}\biggr|_0 &=& \frac{\de^2S_R}{\de \alpha_{mn}^I\de \alpha_{ij}^R}\biggr|_0 = 0\\
\frac{\de^2S_R}{\de \beta_{mn}^R\de \beta_{ij}^R}\biggr|_0 &=& -\frac{\de^2S_R}{\de \beta_{mn}^I\de \beta_{ij}^I}\biggr|_0 = A_+\delta_{mi}\delta_{nj}\\
\frac{\de^2S_R}{\de \beta_{mn}^R\de \beta_{ij}^I}\biggr|_0 &=& \frac{\de^2S_R}{\de \beta_{mn}^I\de \beta_{ij}^R}\biggr|_0 = 0\\
\frac{\de^2S_R}{\de a_{mn}^R\de b_{ij}^R}\biggr|_0 &=& \frac{\de^2S_R}{\de a_{mn}^I\de b_{ij}^I}\biggr|_0 = \frac{\de^2S_R}{\de a_{mn}^R\de b_{ij}^I}\biggr|_0 = \frac{\de^2S_R}{\de a_{mn}^I\de b_{ij}^R}\biggr|_0 = 0\\
\frac{\de^2S_R}{\de a_{mn}^R\de \alpha_{ij}^R}\biggr|_0 &=& \frac{\de^2S_R}{\de a_{mn}^I\de \alpha_{ij}^I}\biggr|_0 = 0\\
\frac{\de^2S_R}{\de a_{mn}^R\de \alpha_{ij}^I}\biggr|_0 &=& \frac{\de^2S_R}{\de a_{mn}^I\de \alpha_{ij}^R}\biggr|_0 = B\delta_{mi}\delta_{nj}\\
\frac{\de^2S_R}{\de a_{mn}^R\de \beta_{ij}^R}\biggr|_0 &=& \frac{\de^2S_R}{\de a_{mn}^I\de \beta_{ij}^I}\biggr|_0 = \frac{\de^2S_R}{\de a_{mn}^R\de \beta_{ij}^I}\biggr|_0 = \frac{\de^2S_R}{\de a_{mn}^I\de \beta_{ij}^R}\biggr|_0 = 0\\
\frac{\de^2S_R}{\de b_{mn}^R\de \alpha_{ij}^R}\biggr|_0 &=& \frac{\de^2S_R}{\de b_{mn}^I\de \alpha_{ij}^I}\biggr|_0 = \frac{\de^2S_R}{\de b_{mn}^R\de \alpha_{ij}^I}\biggr|_0 = \frac{\de^2S_R}{\de b_{mn}^I\de \alpha_{ij}^R}\biggr|_0 = 0\\
\frac{\de^2S_R}{\de b_{mn}^R\de \beta_{ij}^R}\biggr|_0 &=& \frac{\de^2S_R}{\de b_{mn}^I\de \beta_{ij}^I}\biggr|_0 = 0\\
\frac{\de^2S_R}{\de b_{mn}^R\de \beta_{ij}^I}\biggr|_0 &=& \frac{\de^2S_R}{\de b_{mn}^I\de \beta_{ij}^R}\biggr|_0 = B\delta_{mi}\delta_{nj}\\
\frac{\de^2S_R}{\de \alpha_{mn}^R\de \beta_{ij}^R}\biggr|_0 &=& \frac{\de^2S_R}{\de \alpha_{mn}^I\de \beta_{ij}^I}\biggr|_0 = \frac{\de^2S_R}{\de \alpha_{mn}^R\de \beta_{ij}^I}\biggr|_0 = \frac{\de^2S_R}{\de \alpha_{mn}^I\de \beta_{ij}^R}\biggr|_0 = 0
\end{eqnarray*}

where use of the Cauchy-Riemann equations has been made 
(the other derivatives are trivially related to these by 
Schwarz theorem, \eg 
$\frac{\de^2S_R}{\de a_{mn}^R\de a_{ij}^R}=
\frac{\de^2S_R}{\de a_{ij}^R\de a_{mn}^R}$) 
and the coefficients are given by

$$
A_-=2\left(N-N_f\frac{\cosh^2\mu}{m^2}\right) \;\;\;\;\;\;\;
A_+=2\left(N+N_f\frac{\sinh^2\mu}{m^2}\right) \;\;\;\;\;\;\;
B=-2N_f\frac{\cosh\mu\sinh\mu}{m^2}
$$

The hessian for $N=1$ is (with the conventional choice of ordering: $a^R,b^R,\alpha^R,\beta^R,a^I,b^I,\alpha^I,\beta^I$)

$$
H^{\left(1\right)}=
\begin{pmatrix}
A_- & 0 & 0 & 0 & 0 & 0 & B & 0\\
0 & A_- & 0 & 0 & 0 & 0 & 0 & B\\
0 & 0 & A_+ & 0 & B & 0 & 0 & 0\\
0 & 0 & 0 & A_+ & 0 & B & 0 & 0\\
0 & 0 & B & 0 & -A_- & 0 & 0 & 0\\
0 & 0 & 0 & B & 0 & -A_- & 0 & 0\\
B & 0 & 0 & 0 & 0 & 0 & -A_+ & 0\\
0 & B & 0 & 0 & 0 & 0 & 0 & -A_+
\end{pmatrix}
$$

As the second derivatives are manifestly diagonal with respect to the indices $i,j,m,n$, the hessian for a generic $N$ is block-diagonal

$$
H^{\left(N\right)}=\bigoplus^NH^{\left(1\right)}
$$

The model thus features a huge degeneracy of eigenvalues, as the spectrum of $N=1$ is repeated $N$ times.
The spectrum for $N=1$ features $4$ positive eigenvalues and $4$ eigenvalues opposite in sign (as expected by holomorphicity). We are interested in the positive part of the spectrum. An explicit computation shows that the distinct positive eigenvalues of $H^{\left(1\right)}$ are actually $2$ and they are

$$
\lambda_\pm=\frac{1}{2m^2}\left|2N_f\cosh\left(2\mu\right)\pm\sqrt{2}\sqrt{8m^4-8N_fm^2+N_f^2+N_f^2\cosh\left(4\mu\right)}\right|
$$

We note in passing that (here and in many other places) we 
could have written formulas in the complex notation of Takagi 
factorization theorem (see \cite{Kikukawa}), which we decided not
to employ here and in all the paper to stick to a completely real 
notation.

\section{Computing $\Delta_{\hat{n}}(t)$}
\label{sec:DeltaCOMP}

We want to compute $\Delta_{\hat{n}}(t)$, which is defined in 
(\ref{eq:similFP}). The main point is that we are ascending along a
given flow, and while doing that we are transporting along the flow
also the basis vectors, \ie we are integrating (\ref{eq:VecTransport})
as well. Near the critical point the (\ref{eq:asymptSOL})
hold, in which the parametrization 
$\Phi\in\mcal{J}_\sigma\leftrightarrow\left(\hat{n},t\right)$ is manifest. 
Given a reference point $t_0\ll1$, the (\ref{eq:asymptSOL}) can be
regarded as initial conditions for the flow associated to $\hat{n}$. 
Near a generic point, under infinitesimal variations of $t$ and 
$\hat{n}$, the variation of the point $\delta\Phi$ is given by
$$
\delta\Phi=\delta\Phi\left(\hat{n},t\right)=\sum_{i=1}^n V^{\left(i\right)}\left(t\right)\delta c^{\left(i\right)}
$$
This is so because $\delta\Phi$ is itself a vector belonging 
to the tangent space $T_\Phi\mcal{J}_\sigma$. 
The (constant) coefficients $\delta c^{\left(i\right)}$ can be worked 
out from the asymptotic form of $\Phi\left(t\right)$ near the critical 
point

\beqn
\nonumber\delta\Phi\approx\delta\left(\phi_\sigma+\sum_{i=1}^n
v^{\left(i\right)} e^{\lambda_i t}n_i\right)=\sum_{i=1}^n v^{\left(i\right)}
\left(\sum_{j=1}^n\delta n_j\frac{\de}{\de n_j}+
\delta t\frac{\de}{\de t}\right)e^{\lambda_i t}n_i=\\
\nonumber=\sum_{i=1}^n v^{\left(i\right)}e^{\lambda_i t}
\left(\delta n_i+\lambda_i n_i\delta t\right)
\approx\sum_{i=1}^n V^{\left(i\right)}\left(t\right)
\left(\delta n_i+\lambda_i n_i\delta t\right)
\eeqn

from which it follows

$$
\delta c^{\left(i\right)}=\delta n_i+\lambda_i n_i\delta t
$$
Being $\delta\Phi$ a vector of $T_\Phi\mcal{J}_\sigma$, 
we can write it as a decomposition on the (orthonormal) 
$u$-basis 

$$
\delta\Phi=\sum_{i=1}^n u^{\left(i\right)}\delta y_i
$$

and from this we have

$$
\delta y_i=\sum_{j=1}^{2n} u^{\left(i\right)}_j\delta\phi_j
$$

Let us now consider the terms $\frac{\delta y_i}{\delta\star}$ 
appearing in $\Delta_{\hat{n}}\left(t\right)$, 
where $\star$ is either $t$ or $n_j$. For these we have

\beqn
\nonumber\frac{\delta y_i}{\delta\star}=
\sum_{j=1}^{2n}u^{\left(i\right)}_j\frac{\delta\phi_j}{\delta\star}=
\sum_{j=1}^{2n}u^{\left(i\right)}_j\sum_{k=1}^n V^{\left(k\right)}_j
\frac{\delta c^{\left(k\right)}}{\delta\star}=\sum_{j=1}^{2n}u^{\left(i\right)}_j
\sum_{k=1}^n\sum_{l=1}^n u^{\left(l\right)}_j 
E_{lk}\frac{\delta c^{\left(k\right)}}{\delta\star}=\\
\nonumber=\sum_{k=1}^n\sum_{l=1}^n E_{lk}\frac{\delta c^{\left(k\right)}}
{\delta\star}\sum_{j=1}^{2n}u^{\left(i\right)}_j u^{\left(l\right)}_j=
\sum_{k=1}^n\sum_{l=1}^n E_{lk}\frac{\delta c^{\left(k\right)}}
{\delta\star}\delta_{il}=\sum_{k=1}^n E_{ik}
\frac{\delta c^{\left(k\right)}}{\delta\star}
\eeqn

Now we make use of the explicit form of $\delta c^{\left(k\right)}$, 
which gives $\frac{\delta c^{\left(k\right)}}{\delta t}=\lambda_k n_k$ 
and $\frac{\delta c^{\left(k\right)}}{\delta n_j}=\delta_{kj}$, from 
which one can easily derive the (\ref{eq:DLTentries}).

\subsection{$\Delta_{\hat{n}}(t)$ in the gaussian approximation}
\label{sec:DeltaCOMPgss}

We can compute $\Delta_{\hat{n}}\left(t\right)$ for the gaussian case 
(purely quadratic action), where the asymptotic form for the 
SA and parallel-transport equations is correct 
arbitrarily far away from the critical point. 
In that case the entries of 
$\Delta_{\hat{n}}\left(t\right)$ are ($E=\mbb{1}_{n\times n}$)

\beqn
\nonumber\frac{\delta y_i}{\delta t} &=& \lambda_i n_i e^{\lambda_i t}\\
\nonumber\frac{\delta y_i}{\delta n_j} &=& e^{\lambda_i t}\delta_{ij}
\eeqn

and the determinant is

$$
\Delta_{\hat{n}}\left(t\right)=2\left(\sum_{i=1}^n\lambda_i n_i^2\right)
e^{\left(\sum\limits_{i=1}^n\lambda_i\right)t}
$$

For the gaussian action 
$S_R(\hat n,t) = S_R(\phi_\sigma) + \frac{1}{2} \sum_{k=1}^{n}
\lambda_k n_k^2 e^{2 \lambda_k t}$, so that collecting everything
we can write an expression for the weight itself 
$$
\mcal{Z}^{\left(\sigma\right)}_{\hat{n}} = 
2\left(\sum_{i=1}^n\lambda_i n_i^2\right) 
e^{-S_R(\phi_\sigma)} \;
\int_{-\infty}^{\infty} dt \; e^{\sum_{i=1}^n
  \lambda_i t} \;  e^{-\frac{1}{2}\sum_{i=1}^n\lambda_i n_i^2 e^{2
    \lambda_i t}}.
$$
From this expression it is easy to understand that the more the 
eigenvalues differ from each other (which in our case happens for 
low values of the mass parameter), 
the more various 
$\mcal{Z}^{\left(\sigma\right)}_{\hat{n}}$ can differ.

\subsection{A useful consistency relation}

Since in the end we are performing quite a lot of computations, it is
useful to have a consistency relation to be checked while ascending 
along the flow. The gradient of the action is yet another vector 
belonging to the tangent space $T_\Phi\mcal{J}_\sigma$. 
Let us write the decomposition

$$
\nabla_\Phi S^R=\sum_{i=1}^n V^{\left(i\right)}\left(t\right)g^{\left(i\right)}
$$

the coefficients $g^{\left(i\right)}$ can be found with the aid of the 
asymptotic form of the action

\beqn
\nonumber\nabla_\Phi S^R\approx\nabla_\Phi\left(S^R\left(\phi_\sigma\right)
+\frac{1}{2}\Phi^T H\Phi\right)=H\Phi\approx H
\sum_{i=1}^n v^{\left(i\right)}n_i e^{\lambda_i t}=\\
\nonumber=\sum_{i=1}^n v^{\left(i\right)}e^{\lambda_i t}n_i\lambda_i
\approx\sum_{i=1}^n V^{\left(i\right)}\left(t\right)n_i\lambda_i
\eeqn

where we have used the symmetry of the hessian and the fact that 
$Hv^{\left(i\right)}=\lambda_iv^{\left(i\right)}$. 
We have found that $g^{\left(i\right)}=n_i\lambda_i$, so while integrating 
the flow equations, we can keep checked the norm

$$
\left|\nabla_\Phi S^R-\sum_{i=1}^n V^{\left(i\right)}\left(t\right)n_i\lambda_i\right|
$$

and make sure that it is small with respect to the size of the system.


\end{document}